\documentclass{aastex631}

\shorttitle{Urca Pairs in Magnetic Fields}
\shortauthors{Famiano et al.}
\graphicspath{{./}{figures/}}

\begin{document}

\title{Evolution of Urca Pairs in the Crusts of Highly Magnetized Neutron Stars}

\author[0000-0003-2305-9091]{Michael A. Famiano}
\email{michael.famiano@wmich.edu}
\affiliation{Western Michigan University,
Kalamazoo, MI 49008-5252 USA}
\affiliation{National Astronomical Observatory of Japan,
2-21-1 Osawa, Mitaka, Tokyo 181-8588 Japan}
\affiliation{Joint Institute for Nuclear Astrophysics -
Center for the Evolution of the Elements, USA}
\author[0000-0002-3164-9131]{Grant Mathews}
\email{gmathews@nd.edu}
\affiliation{Joint Institute for Nuclear Astrophysics -
Center for the Evolution of the Elements, USA}
\affiliation{Center for Astrophysics, Department of Physics and Astronomy, University of Notre Dame, Notre Dame, IN 46556, USA}

\author[0000-0002-2999-0111]{A. Baha Balantekin}
\email{baha@physics.wisc.edu}
\affiliation{National Astronomical Observatory of Japan,
2-21-1 Osawa, Mitaka, Tokyo 181-8588 Japan}
\affiliation{Department of Physics, University of Wisconsin-Madison, Madison, Wisconsin 53706 USA}

\author[0000-0002-8619-359X]{Toshitaka Kajino}
\email{kajino@buaa.edu.cn}
\affiliation{School of Physics, Beihang University, 37 Xueyuan Road, Haidian-qu, Beijing 100083, China}
\affiliation{National Astronomical Observatory of Japan,
2-21-1 Osawa, Mitaka, Tokyo 181-8588 Japan}
\affiliation{Graduate School of Science, The University of Tokyo, 7-3-1 Hongo, Bunkyo-ku, Tokyo, 113-0033 Japan}

\author[0000-0003-3083-6565]{Motohiko Kusakabe}
\email{kusakabe@buaa.edu.cn}
\affiliation{School of Physics, Beihang University, 37 Xueyuan Road, Haidian-qu, Beijing 100083, China}
\author[0000-0003-2595-1657]{Kanji Mori}
\email{kanji.mori@fukuoka-u.ac.jp}
\affiliation{Research Institute of Stellar Explosive Phenomena, Fukuoka University,\\
8-19-1 Nanakuma, Jonan-ku, Fukuoka-shi, Fukuoka 814-0180, Japan}

\begin{abstract}
We report on the effects of strong magnetic fields on neutrino emission in the modified Urca process.  We show that the effect of Landau levels on the various Urca pairs affects the neutrino emission spectrum and leads to an angular asymmetry in the neutrino emission.
For low magnetic fields the Landau levels have almost no effect on the cooling.  However, as the field strength increases,  the electron chemical potential increases resulting in a lower density at which Urca pairs can exist.
For intermediate field strength there is an interesting interference between the Landau level distribution and the Fermi distribution.  For high enough field strength, the entire electron energy spectrum is eventually confined to single Landau level producing dramatic  spikes in the emission spectrum.  
\end{abstract}


\section{Introduction}
It is by now widely accepted that  soft gamma-ray repeaters (SGRs)
and anomalous X-ray pulsars (AXPs)  correspond to a class of neutron stars known as magnetars.  These objects are warm, isolated, and slowly rotating neutron stars of age $\sim 10^5$ yr with unusually strong surface magnetic fields. Indeed, both
pulsars and magnetars have  strong magnetic fields  at their surface that can be  as large as $10^{12}$ to $10^{16}$ G \citep{Kouveliotou98, Turolla15}.   Moreover,  the interior  field can be as large as the equipartition limit of order $10^{18}$ G  \citep{Lai91,Chanmugam92}.

Among possible explanations for the cooling of magnetars, %
decay processes leading to neutrino or anti-neutrino emission 
such as the direct Urca (DU) process ($n \rightarrow p + e^{-} + {\bar \nu}_e$,
$p + e^{-}  \rightarrow n + \nu_e$),
the modified Urca (MU) process ($n  + N \rightarrow p + e^{-} + N^{\prime} + 
{\bar \nu}_e$,
$p + e^{-} + N \rightarrow n  + N^{\prime} + {\nu}_e$ )
 \citep{haensel94,yakovlev95,yakovlev01},
or the neutrino-pair emission process 
($N_1+ N_2 \rightarrow N_1^{\prime} + N_2^{\prime} + \nu + {\bar \nu}$) 
 are possible.

Early in the cooling process the DU process is a  viable candidate
to explain the rapid cooling of NSs \citep{Boguta81,LPPH91,Maruyama}.
However, in this paper we are more concerned with the MU process.  
The DU process takes place at  high temperature and density and for certain ratios of constituents.  
However, under less extreme conditions when the DU process is diminished,  the MU process can continue to occur.

In particular, the Urca cycles of electron capture and $\beta$ decay on pair nuclei at a depth of about 150 m in neutron stars were shown  to occur  in \cite{schatz14} although this mechanism had been previously discussed in the context of white dwarfs~\citep{tsuruta70}, Type Ia supernovae~\citep{paczynski72,woosley86} and electron-degenerate supernovae~\citep{jones2013}. The Urca cycle operates to cool the outer neutron star
crust by emitting neutrinos while also thermally decoupling the surface layers from the deeper crust. This cooling eliminates the possibility that interior heating produces the unexpectedly short recurrence times of energetic thermonuclear bursts on neutron stars. 
The Urca cycles also indicate that the ignition scenario of superbursts by $^{12}$C+$^{12}$C fusion reaction, whose reaction rate is severely limited in recent theoretical  studies~\citep{mori2019}, would require another heat source because of higher neutrino emissivity for cooling.

In previous studies, however, the effects of possible strong magnetic fields on the modified Urca process have not been considered. 
Here we show that the appearance of Landau levels for electrons  experiencing strong magnetic fields
significantly alters the operation of the various Urca pairs proposed by \cite{schatz14} and
affects the neutrino emission spectrum leading to an angular asymmetry in the neutrino emissivity.
For low magnetic fields the Landau levels have almost no effect on the cooling.  However, as the field strength increases,  the electron chemical potential increases resulting in a lower density at which Urca pairs can exist.
For intermediate field strength there is an interesting interference between the Landau level distribution and the Fermi distribution.  For high enough field strength, eventually the entire electron energy spectrum is confined to single Landau level producing dramatic  spikes in the emission spectrum.  

This paper is organized as follows:  The ingredients of the model for the MU are summarized in Section 2.  The results are presented in Section 3.  Our discussion and conclusions are in Section 4.

\section{The Model}
\subsection{Weak Interaction Rates in External Fields}
In a homogeneous plasma at a given temperature and density, the electron chemical potential will change in the presence of a strong
magnetic field.  This will ultimately change the effectiveness of a particular Urca pair for a given
environment.  If a dipole field is assumed, then the electron chemical potential, which depends on the magnetic
field, will in turn depend on the angular position with respect to the magnetic pole for a constant
density and temperature within the crust. The  electron
number density with the electron transverse momentum components constrained to Landau  levels is \citep{famiano20,grasso01,kawasaki12}:
\begin{eqnarray}
\label{derv}
n_e = n_- - n_+ &= \frac{1}{4\pi^2}\int_{0}^{\infty}d^3p
{\left(
\left[{\exp{\left(\frac{E-\mu}{T}\right)}+1}\right]^{-1} 
- 
\left[{\exp{\left(\frac{E+\mu}{T}\right)}+1}\right]^{-1}
\right)}\\\nonumber
&= \int_{0}^{\infty}dp_xdp_ydp_z
{\left(
\left[{\exp{\left(\frac{E-\mu}{T}\right)}+1}\right]^{-1} 
- 
\left[{\exp{\left(\frac{E+\mu}{T}\right)}+1}\right]^{-1}
\right)}\\\nonumber
&=\frac{eB}{2\pi^2}\sum\limits_{n=0}^{\infty}g_n\int_{0}^{\infty}dp_z
\left(
\left[{\exp{\left(\frac{\sqrt{p_z^2 + m_e^2 + 2n eB}-\mu}{T}\right)}+1}\right]^{-1}
\right.
\\\nonumber
&- 
\left.\left[{\exp{\left(\frac{\sqrt{p_z^2 + m_e^2 + 2n eB}+\mu}{T}\right)}+1}\right]^{-1}
\right)
\end{eqnarray}
where $g_n$ is the degeneracy of individual Landau levels $n$:
\begin{equation}
    g_n = \left\lbrace
    \begin{array}{cl}
        1,  &  n=0\\
        2,  & n>0
    \end{array}
    \right.
    .
\end{equation}
Natural units are used in this manuscript ($k=\hbar=c=1)$.

Weak interaction rates are computed as in \cite{famiano20} and \cite{arcones10}.
In the presence of an external field, the electron energy density  is
reconfigured by the presence of the field such that the 
components of the electron momentum perpendicular to the external magnetic 
field,   $p_\perp^2=p_x^2+p_y^2$, are placed into individual levels, $p_{\perp,n}^2 = neB$.
This quantization results in a shift in the electron chemical potential, $\mu_e$, and ultimately
the weak interaction rates \citep{luo20,grasso01,fassio69}.  At low fields, $B\lesssim B_c$,
 where $B_c\sim {m_e^2}/{e}=4.4\times 10^{13}$ G, the Fermi distribution of the electrons is very similar  to the
distribution in which B=0.  In this region, the spacing between individual Landau levels is small,
and the phase space distributions in Equation (\ref{derv}) between the magnetized plasma and the
non-magnetized plasma are similar \citep{luo20,grasso01,fassio69}:
\begin{equation}
    dn \propto \frac{d^3p}{(2\pi)^3} = \sum\limits_{n=0}^{\infty}\left(2 - \delta_{n0}\right)\frac{eB}{2\pi^2}dp_z.
\end{equation}
From this,
the 
 Fermi-Dirac distribution for the $n^{th}$ Landau level is rewritten:
\begin{equation}
    f_{FD}(E,\mu_e) = \frac{1}{\exp{\left[\frac{\sqrt{E^2 + 2neB}-\mu_e}{T}\right]}+1}.
\end{equation}

Following the prescription of \cite{arcones10} and \cite{famiano20}, the weak interaction rates are 
approximated
as:
\begin{eqnarray}
\label{mag_beta_minus}
    \Gamma_{\beta^-} & = &\kappa \frac{eB}{2} \sum\limits_{n=0}^{N_{max}}\left(2-\delta_{n0}\right)\int\limits_{\omega_\beta}^{Q}\frac{E(Q-E)^2}{\sqrt{E^2 - m_e^2-2neB}}\left(1-f_{FD}(E,\mu_e)\right)\left(1-f_{FD}(Q-E,-\mu_\nu)\right)dE,
    \\
    \label{mag_beta_plus}
    \Gamma_{\beta^+} & = &\kappa \frac{eB}{2} \sum\limits_{n=0}^{N_{max}}\left(2-\delta_{n0}\right)\int\limits_{\omega_\beta}^{-Q}\frac{E(-Q-E)^2}{\sqrt{E^2 - m_e^2-2neB}}\left(1-f_{FD}(E,-\mu_e)\right)\left(1-f_{FD}(-Q-E,-\mu_\nu)\right)dE,
    \\
    \label{mag_beta_EC}
    \Gamma_{EC} & = &\kappa \frac{eB}{2} \sum\limits_{n=0}^{N_{max}}\left(2-\delta_{n0}\right)\int\limits_{\omega_{EC}}^\infty\frac{E(E-Q)^2}{\sqrt{E^2 - m_e^2-2neB}}f_{FD}(E,\mu_e)\left(1-f_{FD}(E-Q,\mu_\nu)\right)dE,
    \\
    \label{mag_beta_PC}
    \Gamma_{PC} & = &\kappa \frac{eB}{2} \sum\limits_{n=0}^{N_{max}}\left(2-\delta_{n0}\right)\int\limits_{\omega_{PC}}^\infty\frac{E(E+Q)^2}{\sqrt{E^2 - m_e^2-2neB}}f_{FD}(E,-\mu_e)\left(1-f_{FD}(E+Q,-\mu_\nu)\right)dE,
\end{eqnarray}
in which the following are defined:
\begin{eqnarray}
    \omega_{EC/PC} &\equiv & \mbox{max}\left[\pm Q,m_e\right],
    \\\nonumber
    \omega_{\beta} &\equiv & \sqrt{m_e^2+2neB},
    \\\nonumber
    N_{max}&\le &\frac{Q^2 - m_e^2}{2eB},
    \\\nonumber
    \kappa & \equiv & \frac{B\ln 2}{K m_e^5},
    \\\nonumber
    B & \equiv & 1+3g_A^2 
    =
    \left\lbrace
    \begin{array}{ll}
      5.76, & \text{nucleons}, \\
      4.6, & \text{nuclei},
    \end{array}
    \right.
    \\\nonumber
    K &\equiv & \frac{2\pi^3\hbar^7\ln 2}{G_V^2 m_e^5} = 6144 \mbox{ s}
\end{eqnarray}
and $Q$ is the nuclear mass difference between  the parent and daughter nucleus.

In this evaluation, we recognize that the above rates are semi-classical approximations in which
nuclear structure and sums over excited states are ignored.  We take this approach to examine
the overall gross effects of the external field.  The rates evaluated above can be adapted to individual
sums over individual transitions from parent to daughter states if desired.  In this work, we simplify
the above to examine ratios of transition rates in an external field to those without a field.

In a charge-neutral plasma at temperature $T$ and electron charge density $\rho Y_e/N_A$, the electron chemical potential is calculated from  Eq.~(\ref{derv}).  
The chemical potential is then used in Eqs.~(\ref{mag_beta_minus}) -- (\ref{mag_beta_PC}) to evaluate
the modified rates.  
\subsection{Urca Pairs}
For an Urca pair to exist \citep{schatz14}, a pair of nuclei, $^AZ$ and $^A(Z+1)$ must be  linked by weak interactions
\begin{eqnarray}
    ^AZ\rightarrow ^A(Z+1)+e^-+\bar{\nu}_e\\\nonumber
    ^A(Z+1)+e^-\rightarrow ^AZ+\nu_e ~~.
\end{eqnarray}
The electron energy distribution in the plasma must also be non-degenerate at energies less than the $\beta^-$ decay Q value and non-zero at energies greater than the electron-capture Q value.  In other words, 
the electron phase space must be simultaneously available for both $\beta^-$ decays and electron capture.
For a plasma in which this condition is met $\mu_e\approx Q_{\beta^-}=-Q_{EC}$.  The finite plasma temperature results in the 
 availability  of electron states at energies below Q and electrons occupying states at energies above Q.  Under these conditions,
the Urca pair will undergo  captures and decays, resulting in enhanced cooling via neutrino
emission.  Captures may occur to low-lying states of the daughter nucleus~\citep{schatz14}.

In the absence of a magnetic field, the electron chemical potential is constrained by the temperature, $T$, and  the charge density $\rho Y_e$ of the plasma.   That is, for a specific temperature, there is only a one-to-one
relationship between electron chemical potential and charge density.   In fact,  there exists a range
of charge densities that define an Urca pair because of the finite plasma
temperature.  The Urca pair is constrained to the region where $Q-kT\lesssim\mu\lesssim Q+kT$.  Thus, 
for a specific temperature, $T$, and electron chemical potential, $\mu$, a range of densities
for which an Urca pair exists can be defined, $\rho Y_e(T,\mu-kT)\lesssim \rho Y_e\lesssim \rho Y_e(T,\mu+kT)$.  It has been  found ~\citep{schatz14,deibel16} that this density range results in a thin layer in which a particular pair can exist.

However, this 
constraining relationship is broken by the introduction of an external magnetic field, $\mu = \mu(T,\rho Y_e, B)$.  This may have multiple effects because:
\begin{itemize}
    \item The electron chemical potential depends on the external field, the density at which
    an Urca pair forms for a particular temperature changes, resulting in a changed emissivity.
    \item Weak rates may change at various field strengths, changing the neutrino emissivity.
    \item For a specific  T, $\rho Y_e$, and B, new Urca pairs may result from changes in the chemical potential.  In addition,  existing Urca pairs may by prohibited as the field changes.
    \item EC rates on exited state nuclei will change with  the magnetic field strength.  Concurrent changes in the 
     $\beta$-decay rates of the daughter nucleus may result in similar rates between exited states
     and the existance of Urca pairs which involve formerly inaccessible excited states in nuclei.
     
     \item For a non-uniform magnetic field on the surface of a highly-magnetized neutron star,
     the location of Urca pairs in the ocean and crust of a neutron star may depend on  the location on the surface of the star.
\end{itemize}

For a specific Urca pair, the emissivity and luminosity can be calculated following the method of 
\cite{deibel16}:
Our model for the neutron star and the emissivity in the modified Urca process are directly
dependent on the weak interaction rates~\citep{tsuruta70,deibel16}:
\begin{equation}
    \epsilon_{\pm}\approx m_e \Gamma_\pm .
\end{equation}
In this work, we define the quantity $\epsilon_{22}$:
\begin{equation}
    \epsilon_{22}\equiv \frac{\left(\epsilon_-+\epsilon_+\right)}{10^{22}}
\end{equation}

The geometric thickness of an Urca layer can be shown to be quite thin.  In this case, 
the luminosity (in units of 10$^{32}$ erg s$^{-1}$ as a function of polar angle, assuming a relativistic correction is:
\begin{equation}
    L_{32}(\theta) \equiv \varepsilon\times2\pi R^2\sin(\theta)\Delta\theta\Delta R/(10^{32} \mbox{erg s}^{-1})
\end{equation}
The radial
thickness of each zone is calculated using the formulation of
\cite{schatz14} and is found to be $\sim$ 1 m.  
\section{Results}
\begin{figure}
    \centering
        \gridline{
        \fig{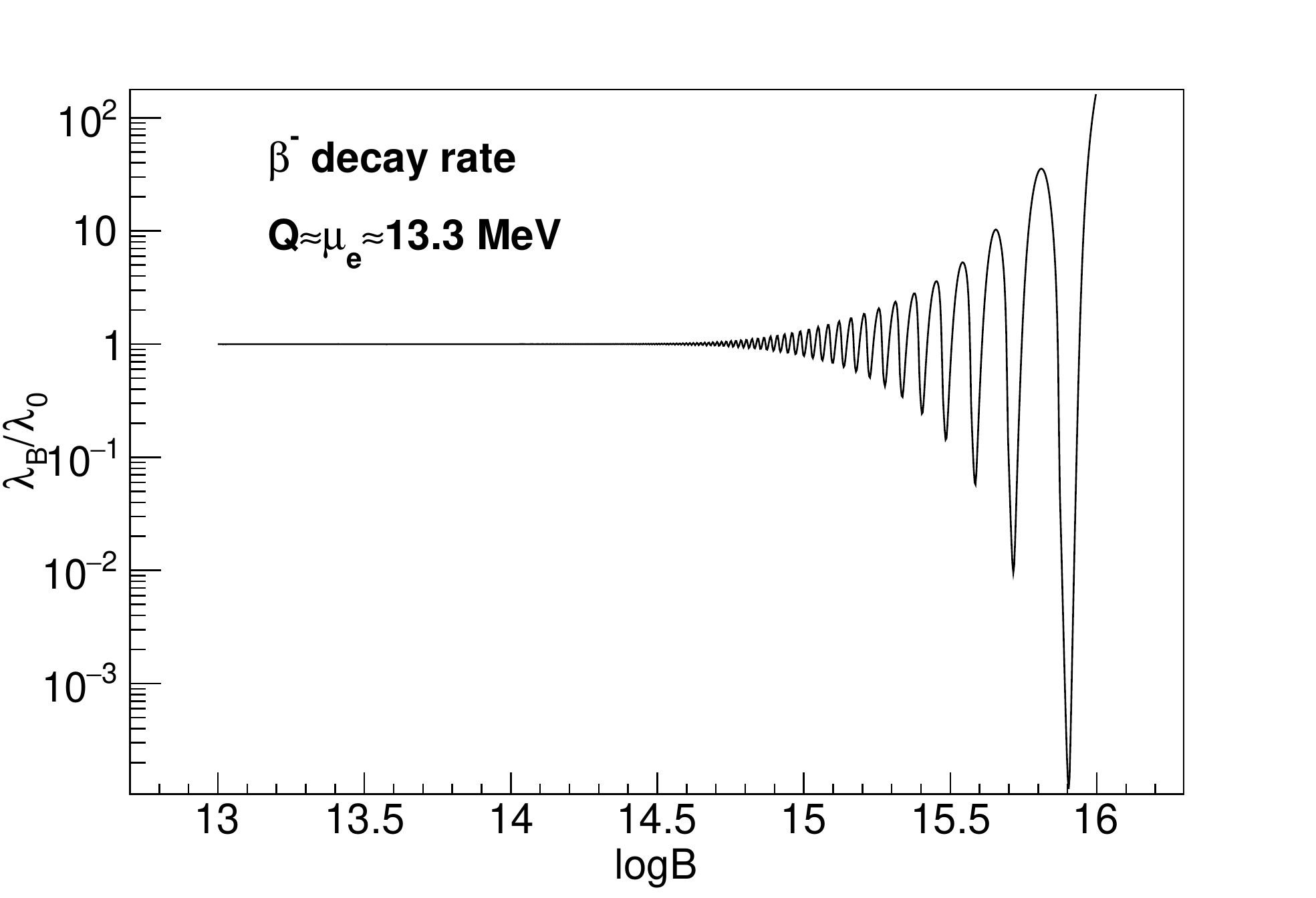}{0.5\textwidth}{(a)}
        \fig{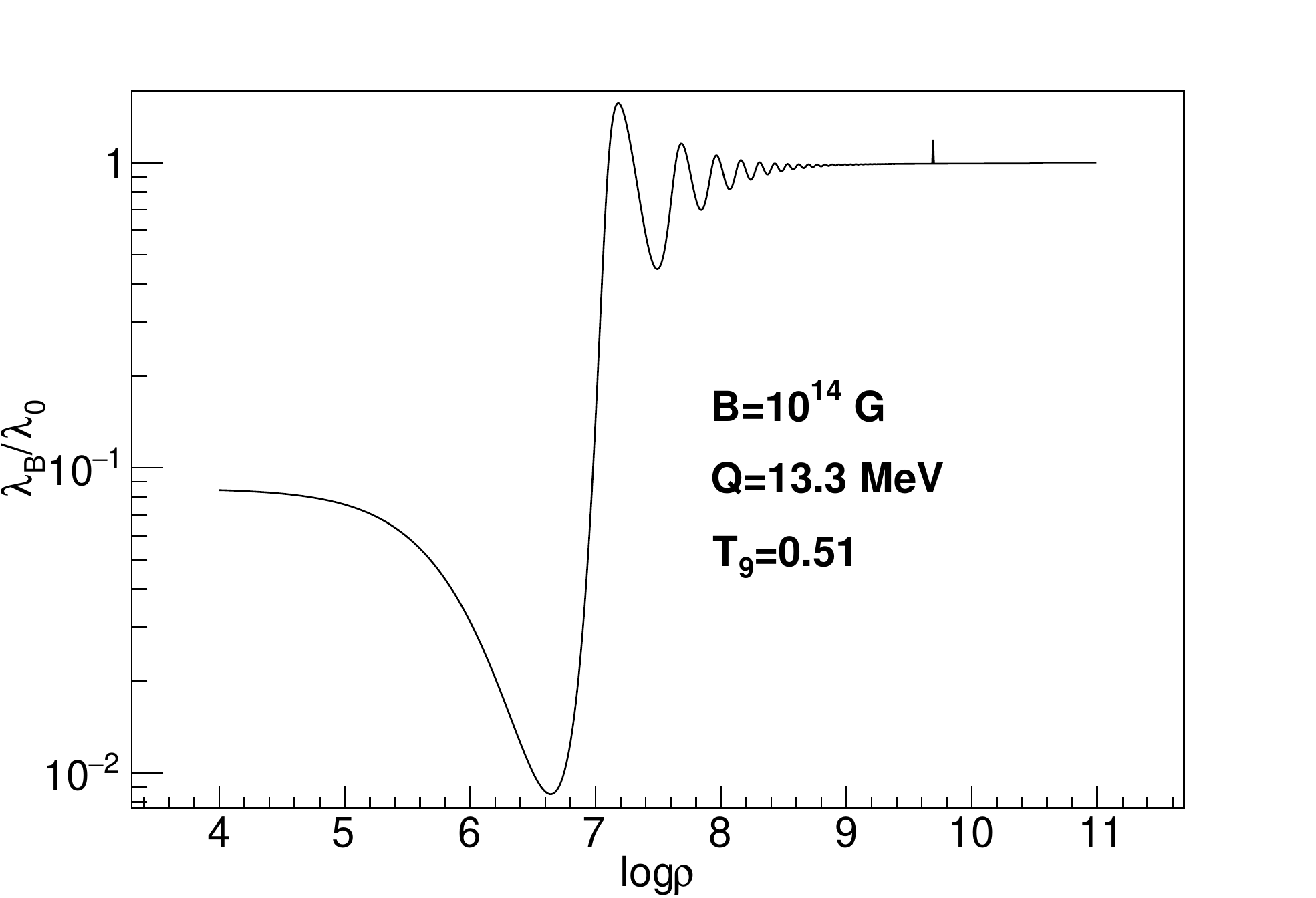}{0.5\textwidth}{(b)}
    }
    \gridline{
        \fig{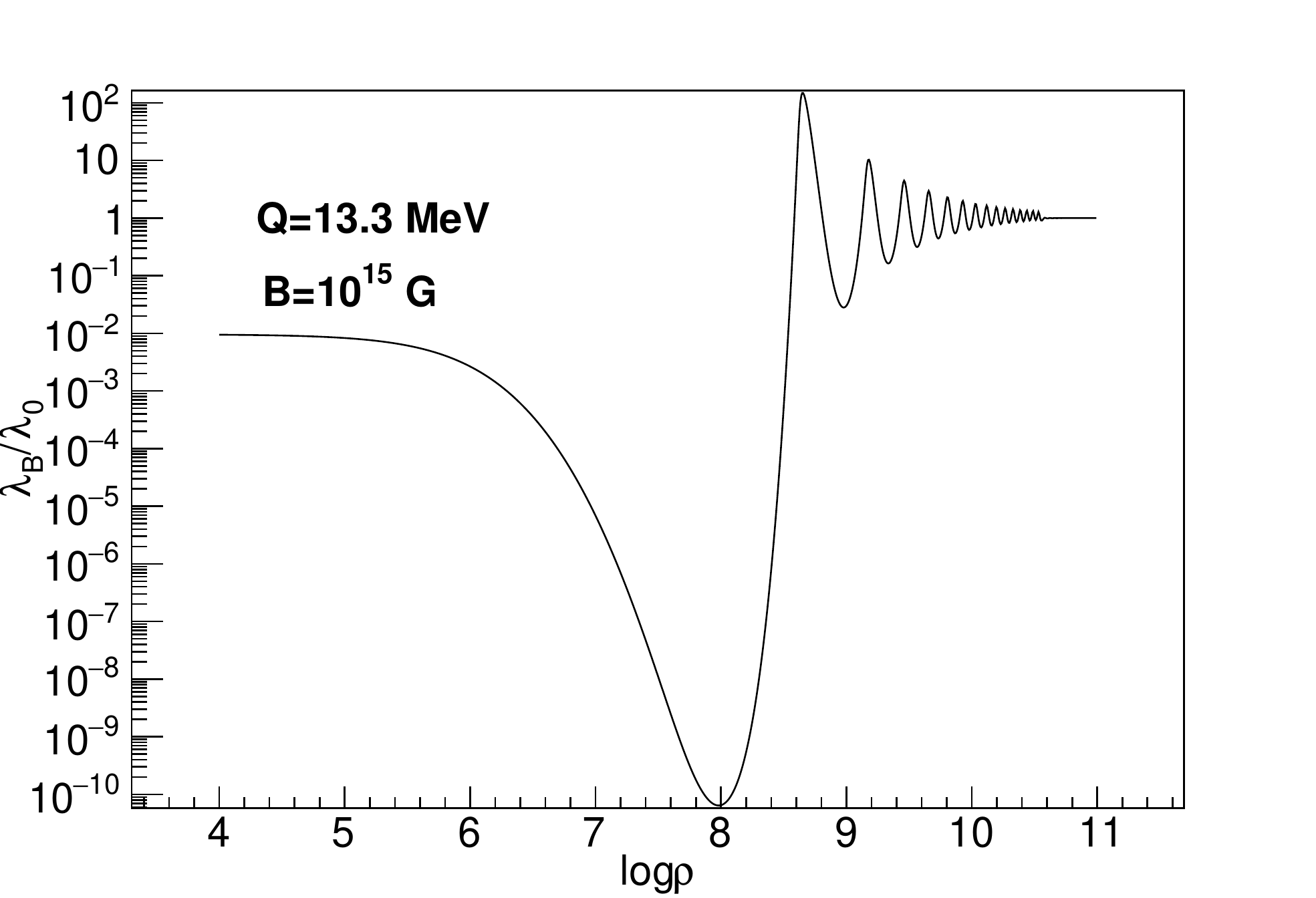}{0.5\textwidth}{(c)}
    }   
    
    \caption{(a)  Ratio of $\beta^-$ decay rates with and without a magnetic field as a function of magnetic field strength} for a temperature and $\rho Y_e$ resulting in an electron chemical potential approximately equal to a Q value of 13.3 MeV.
    (b) Rate ratio as a function of density for $Y_e$=0.5, $T_9$=0.51, and B=10$^{14}$ G.
    (c) Rate ratio as a function of density for $Y_e$=0.5, $T_9$=0.51, and  B=10$^{15}$ G.
    \label{beta_ratios}
\end{figure}

The $\beta^-$ decay rate ratio $\lambda(B)/\lambda(B=0)$ is shown as a function of the magnetic
field in Figure \ref{beta_ratios}(a) for $T_9$=0.51, $Y_e$=0.5, and $\rho$=
4$\times 10^{10}$ g cm$^{-3}$.  The oscillations in the decay rate occurs with a change in
the field as fewer Landau levels contribute to the electron energy spectrum and as Landau levels shift
across the spectrum.

The decay rate ratio as a function of density is shown in Figure \ref{beta_ratios}(b) for $T_9$=0.51
and B=10$^{14}$ G.  While the oscillatory behavior of Figure
\ref{beta_ratios}(a) can be explained by Landau levels shifting into or out of the 
electron energy spectrum as the field changes, in Figure \ref{beta_ratios}(b), the 
oscillatory behavior can be explained by the shift in electron chemical potential as the density increases.
The change in chemical potential results in the electron energy spectrum shifting with
respect to the existing Landau levels.  The same behavior is shown for a field of 10$^{15}$ G in Figure
\ref{beta_ratios}(c).

\begin{figure}
    \centering
        \gridline{
        \fig{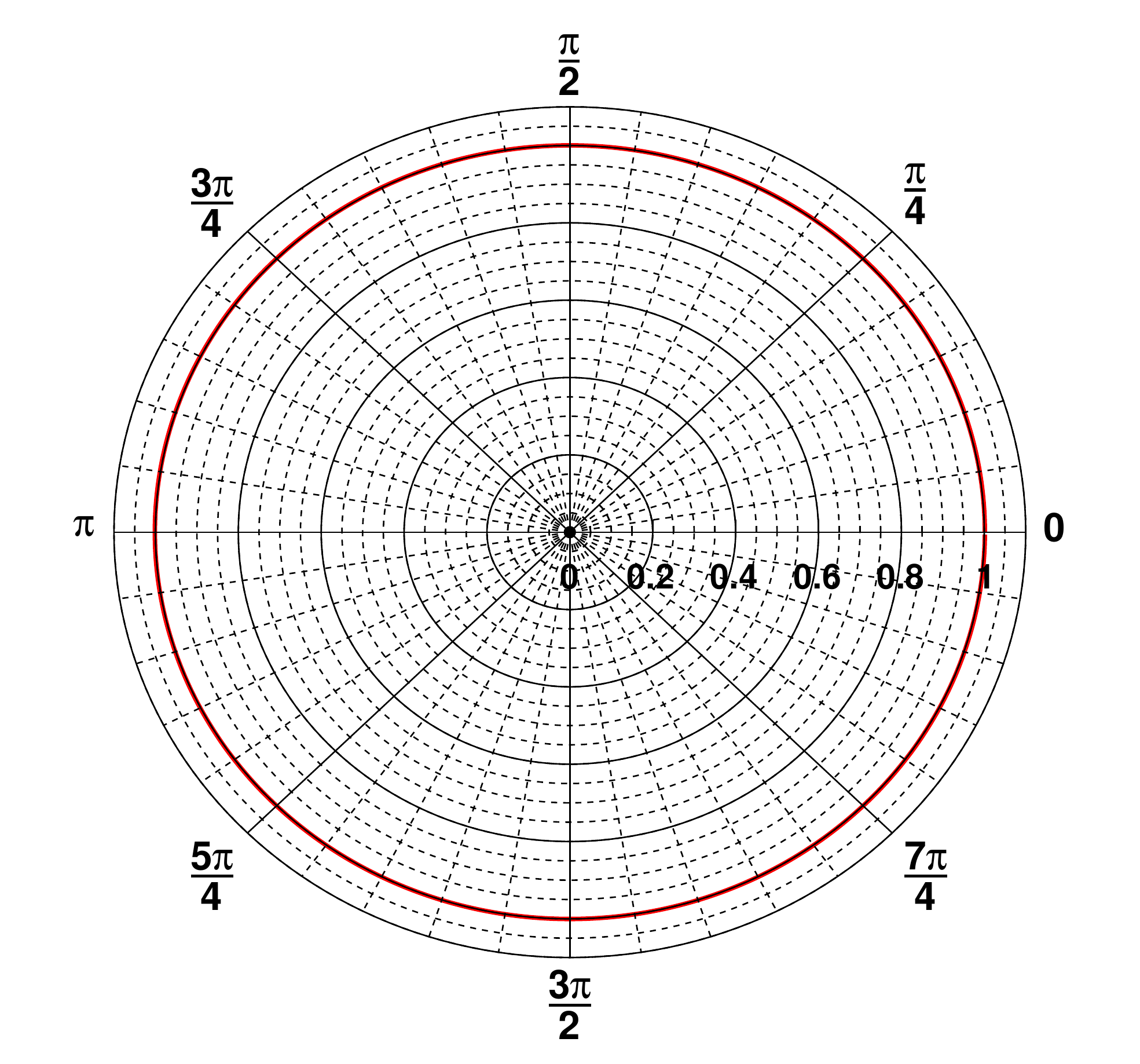}{0.33\textwidth}{(a)}
                \fig{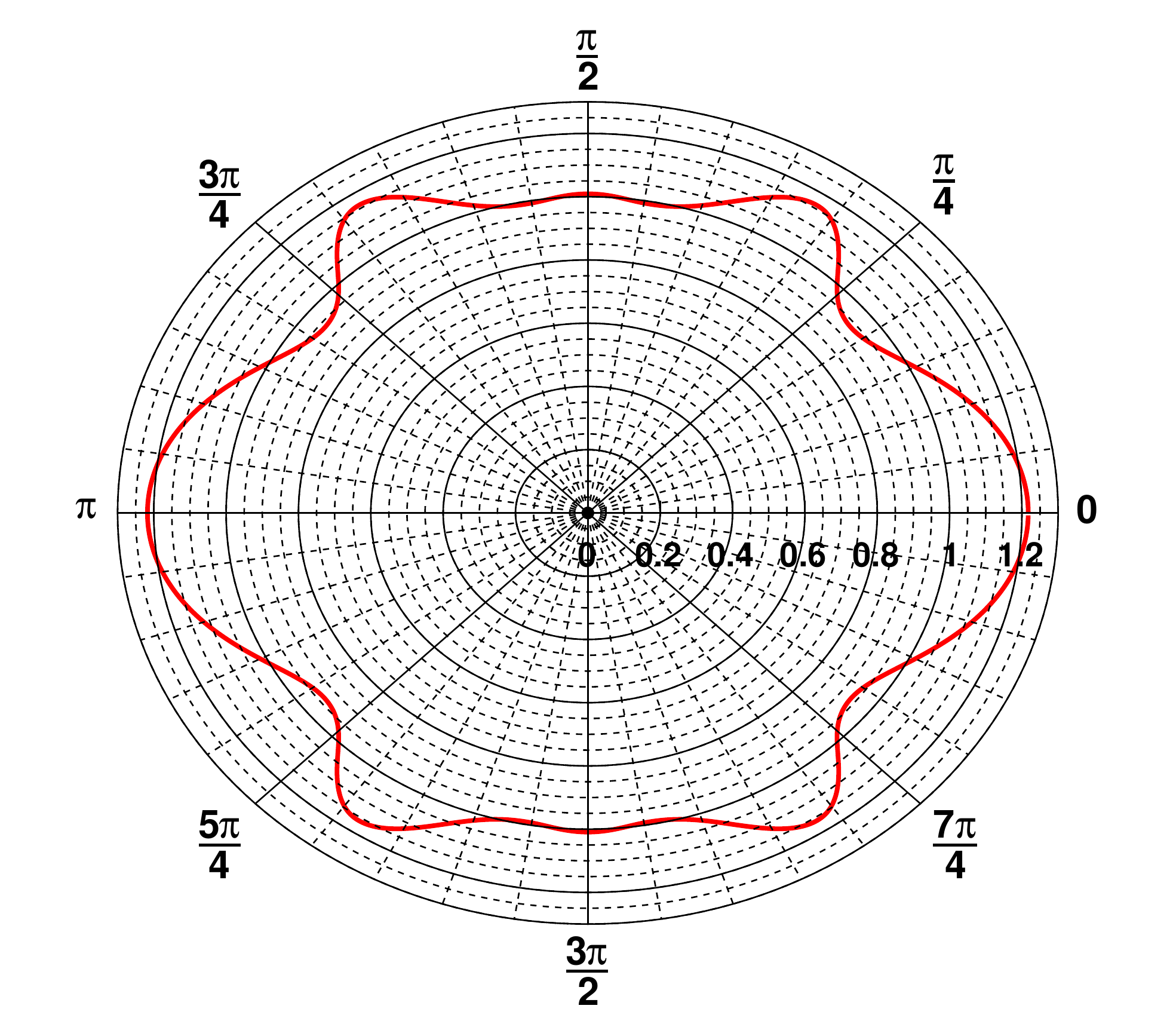}{0.33\textwidth}{(b)}
                        \fig{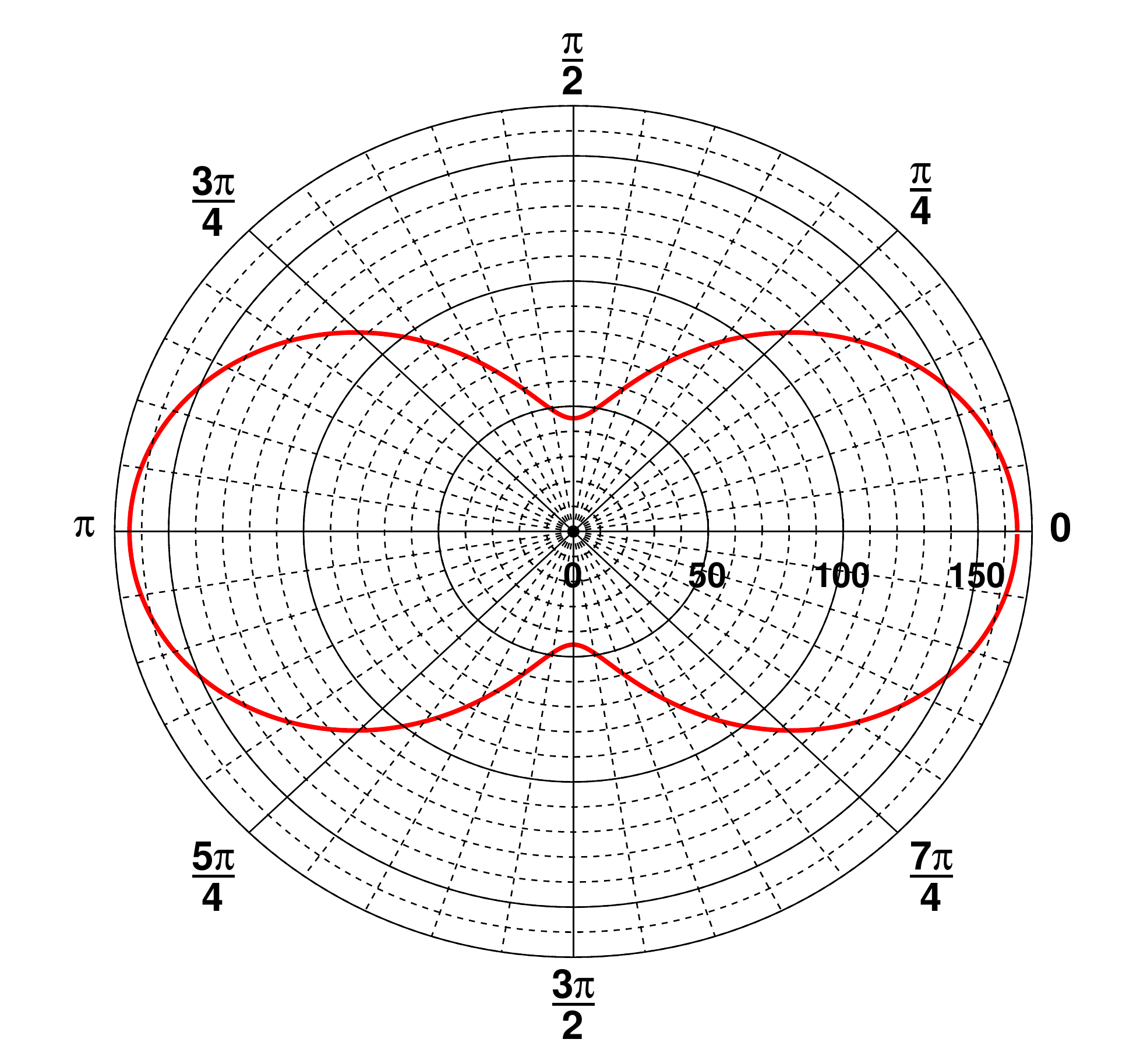}{0.33\textwidth}{(c)}
    }
    \gridline{
        \fig{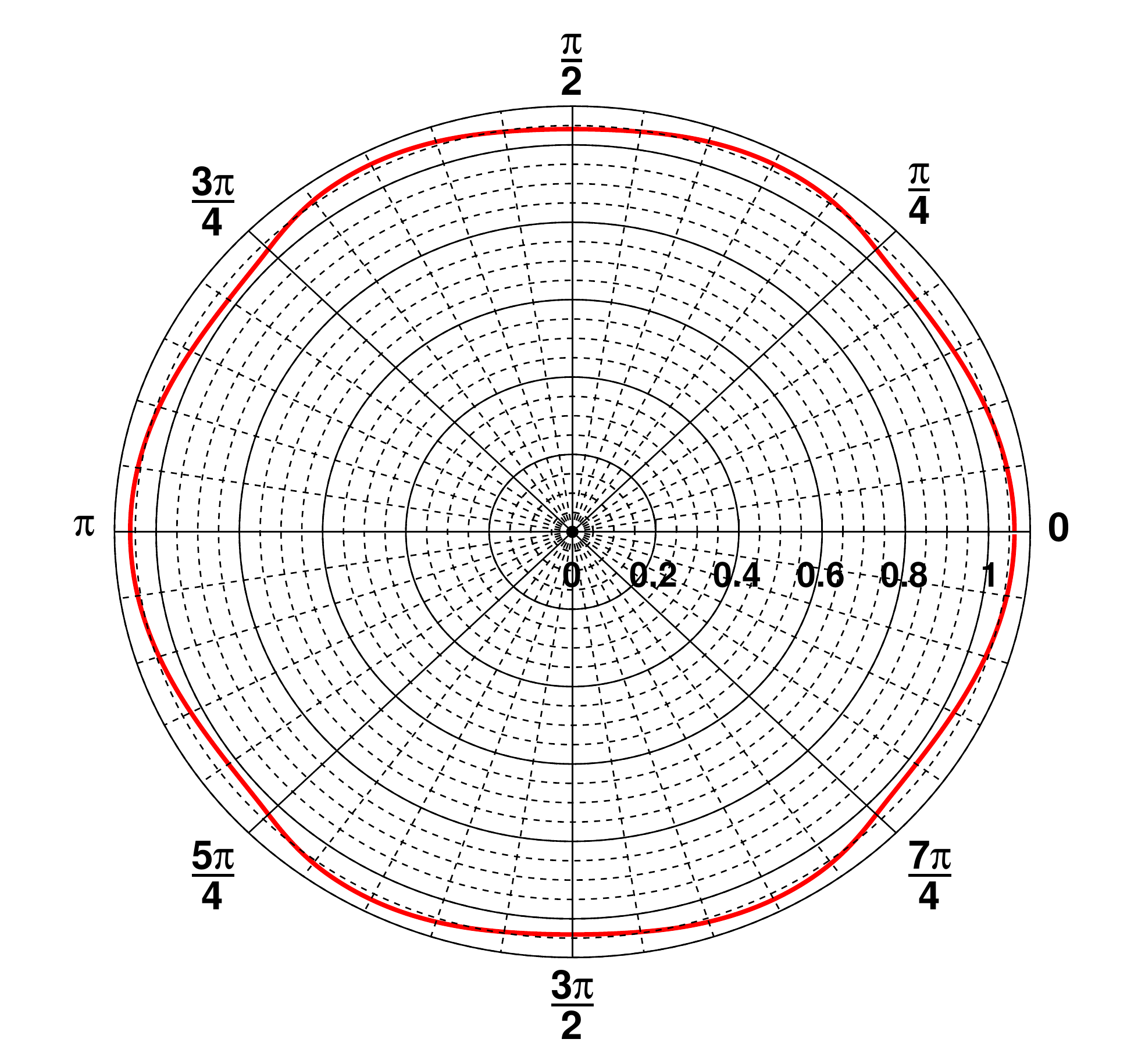}{0.33\textwidth}{(d)}
                \fig{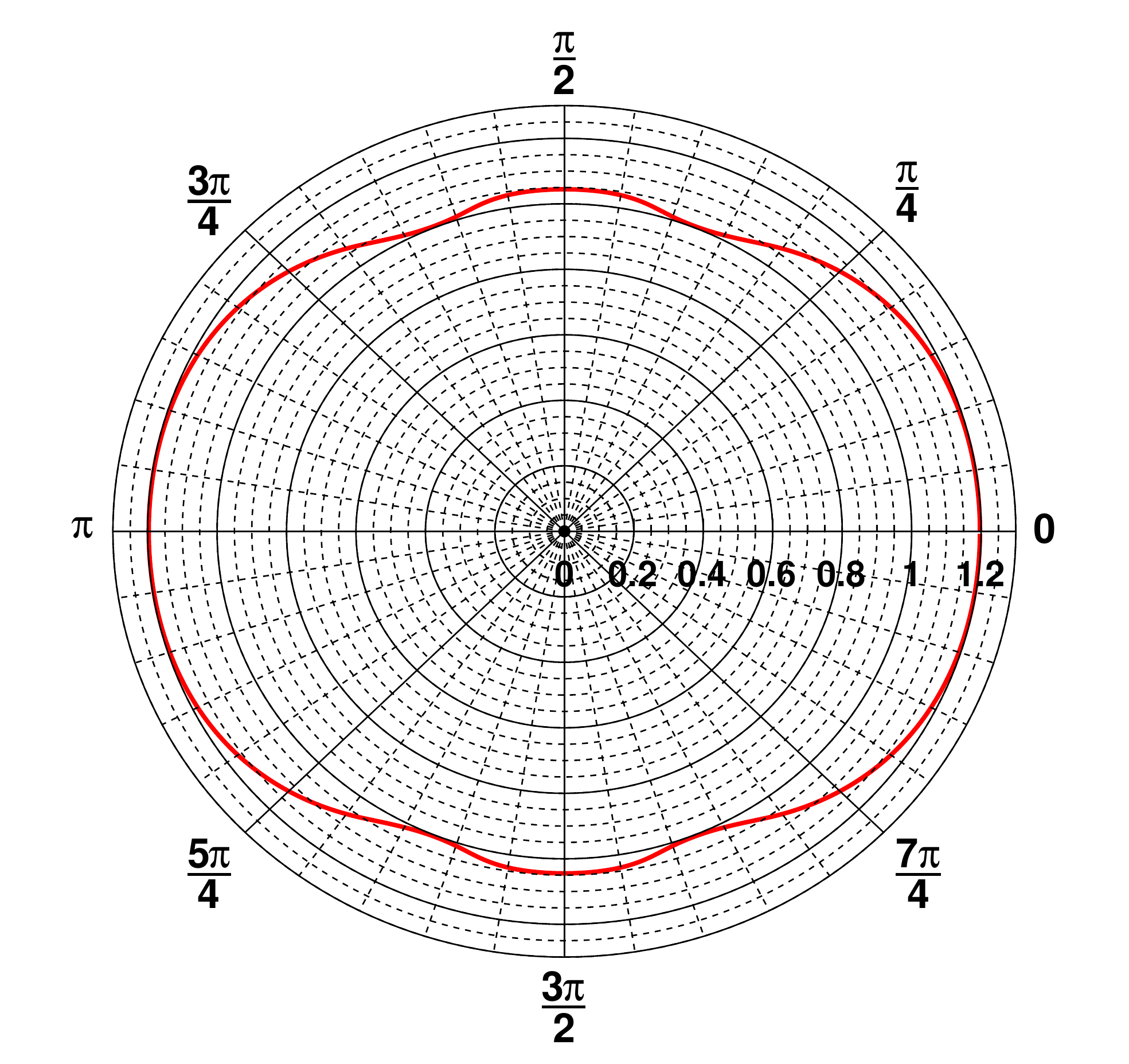}{0.33\textwidth}{(e)}
                        \fig{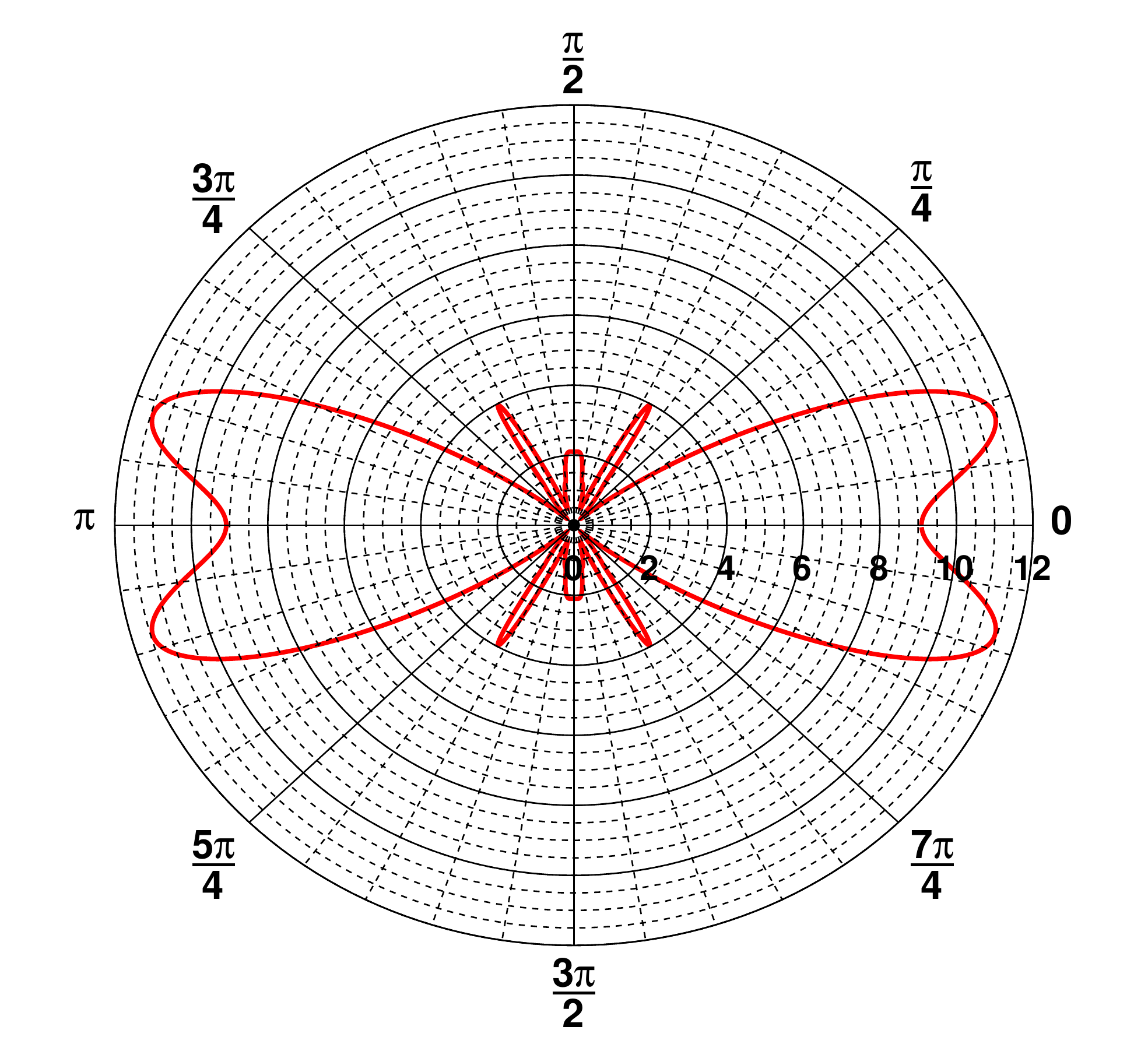}{0.33\textwidth}{(f)}
    }
	\caption{Ratios of $\beta^-$ decay rates $\lambda(B\ne 0)/\lambda(B=0)$ as a function of polar angle, $\theta$ as measured from
		a magnetic pole of the star for $Q_{\beta^-}=$5 MeV.  The top row shows 
		the field dependence at a constant $\rho Y_e$ = 10$^9$ g cm$^{-3}$ for (a) B=10$^{14}$ G, (b) B=10$^{15}$ G, and (c) B=10$^{16}$ G.  The bottom row
		shows the sensitivity to density at a constant B=10$^{15}$ G for 
		$\rho Y_e/10^9$ of (d) 0.3 g cm$^{-3}$, (e) 0.6 g cm$^{-3}$, and (f) 2 g cm$^{-3}$.}
    \label{latitude}
\end{figure}

 Figure \ref{latitude} shows the ratio of electron capture rates $\lambda(B)/\lambda(B=0)$ as a function
of polar angle about the axis of the neutron star for two different values of field (as indicated at
the pole).  The field is assumed to be a dipole field.  Electron capture rates are computed at the stellar 
surface.  For each case, $Q_{EC}$ = 5 MeV.  For a weaker field, more Landau levels are present in the 
electron phase space, and as the field changes
across the surface of the star a larger number of Landau levels are shifted out of the electron
energy spectrum resulting in the larger number of fluctuations in the EC rate.  However, for a larger
field, fewer Landau levels are included in the electron energy spectrum, and as the field shifts across 
the surface of the star, fewer Landau levels shift into or out of the electron energy spectrum.  Thus, there are fewer oscillations in the EC rate across the surface of the star.

\begin{figure}
    \centering
        \gridline{
        \fig{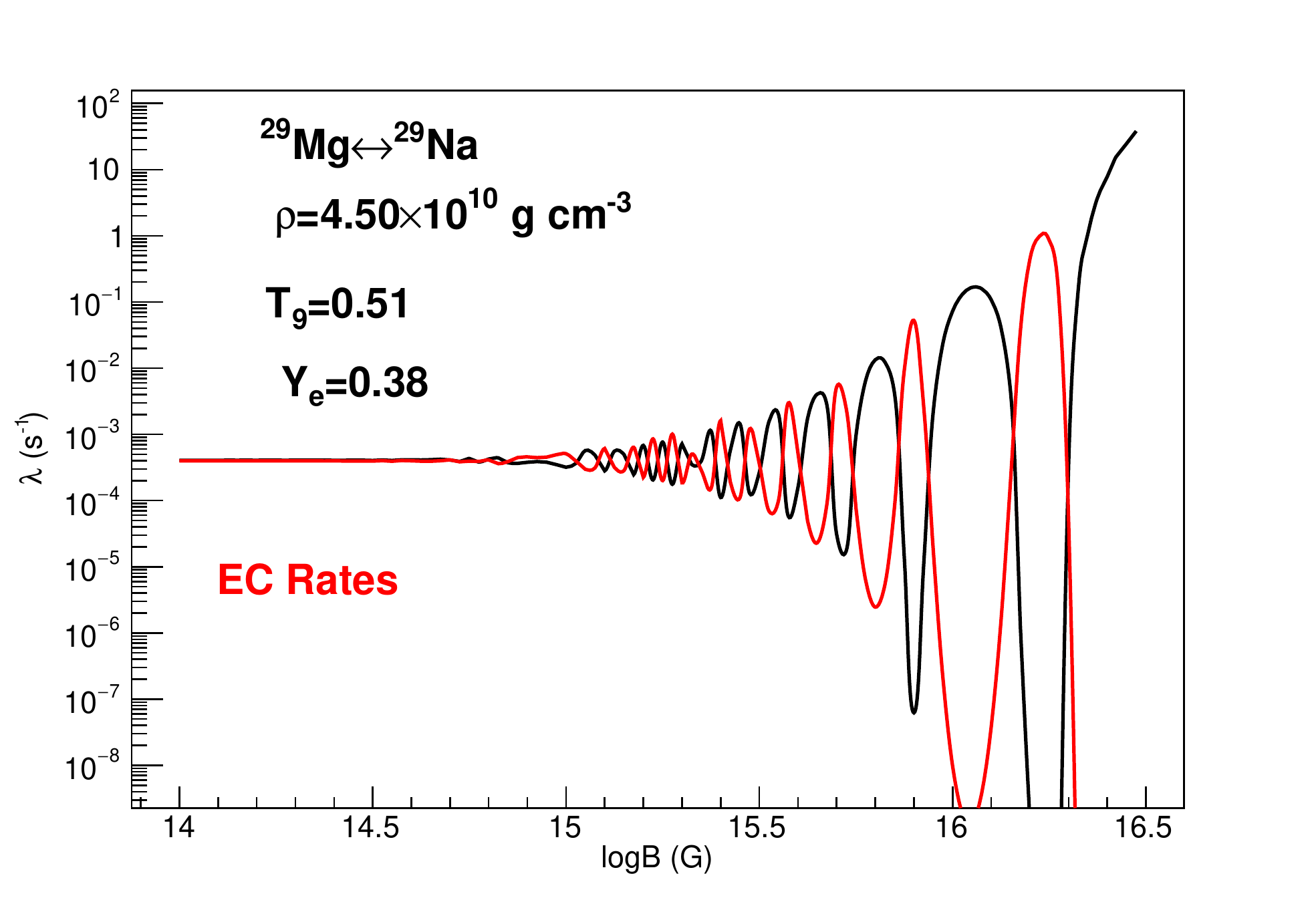}{0.5\textwidth}{29Mg}
        \fig{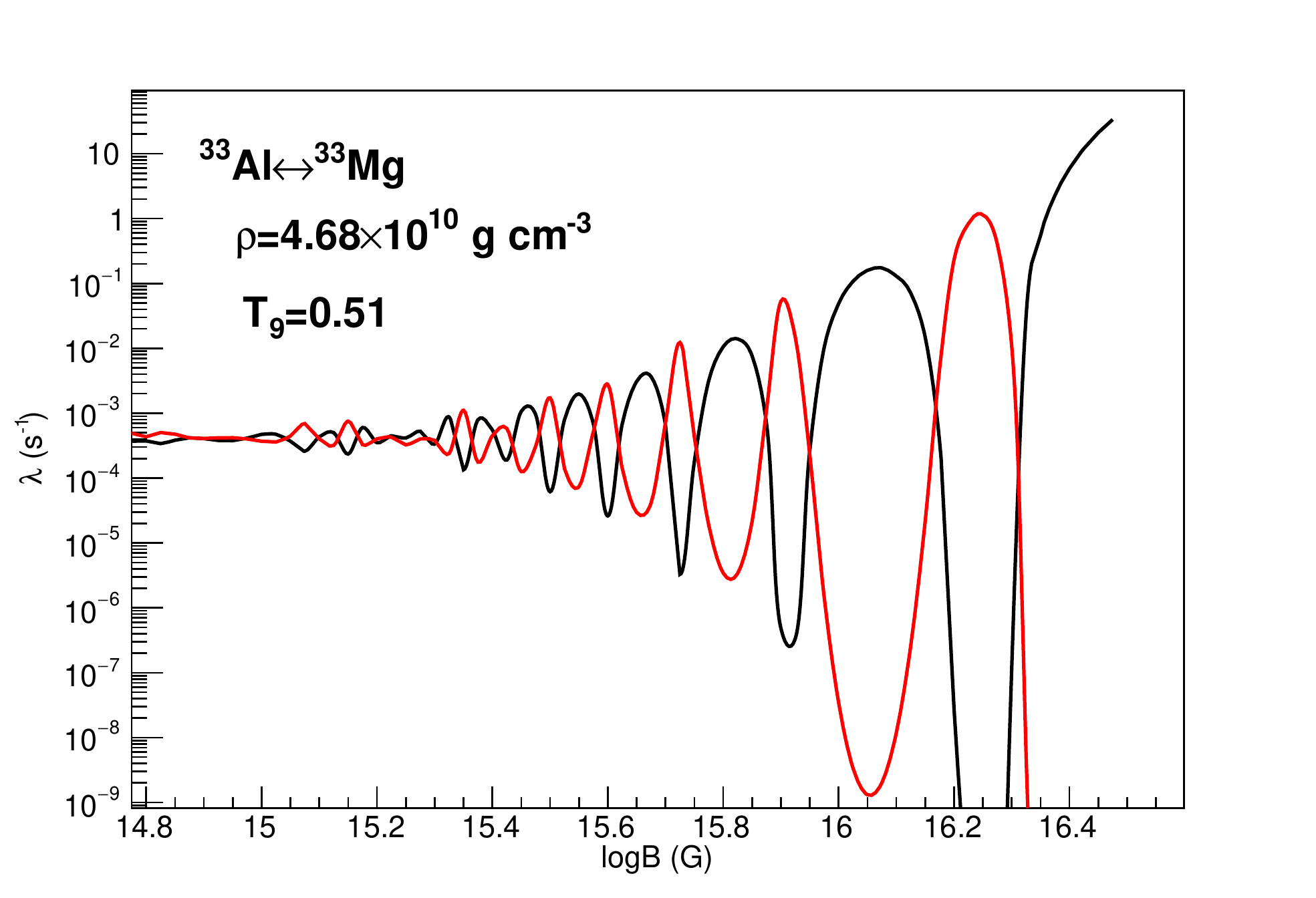}{0.5\textwidth}{33Al}
    }
	\caption{Examples of EC and $\beta^-$ decay rates as a function of magnetic field for
	$^{29}$Mg$\leftrightarrow^{29}$Na (left) and $^{33}$Al$\leftrightarrow ^{33}$Mg (right).
	The red line corresponds to the EC rate, while black lines correspond to the $\beta^-$ rate.}
    \label{mg33_fig}
\end{figure}

In order for efficient Urca pair cooling  to occur, EC and $\beta^-$ decay rates must be similar.  We
thus examine these rates as a function of magnetic field for several Urca pairs.  In
Figure \ref{mg33_fig}, rates are shown as a function of magnetic field for two presumed
Urca pairs~\cite{schatz14}.  In each figure, the black line corresponds to $\beta^-$ decay
rates while the red line corresponds to EC rates.  Optimal Urca pair cooling occurs if the rates
are similar.  However, as the field increases, there can be oscillations in the rates, reducing
the efficiency of the associated pair.  For example, consider  the $^{29}$Mg$\leftrightarrow^{29}$Na pair
in Figure \ref{mg33_fig} at the temperature, density, and electron fraction indicated.
At a field of B$\approx 10^{15.8}$, the EC rate exceeds the $\beta^-$ rate by about five
orders of magnitude.  However, at a field  strength slightly lower than this, the rates intersect on the graph.
  At very high fields, the rates diverge significantly, as only one Landau level
contributes to the electron energy spectrum.  A similar comparison can be made for 
the $^{33}$Al$\leftrightarrow ^{33}$Mg pair in the same figure.

\begin{figure}
    \centering
    \gridline{
        \fig{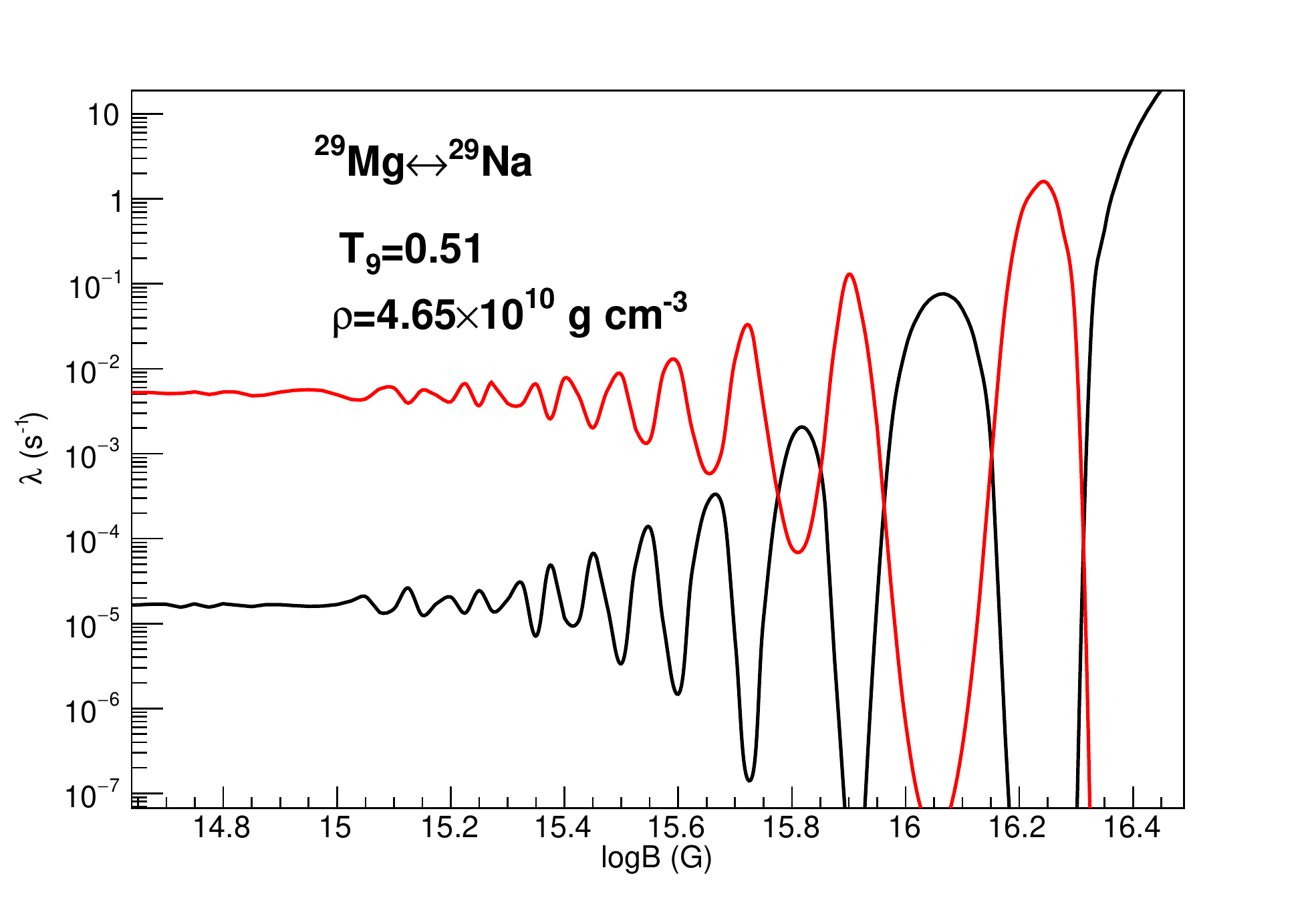}{0.5\textwidth}{29Mg}
        \fig{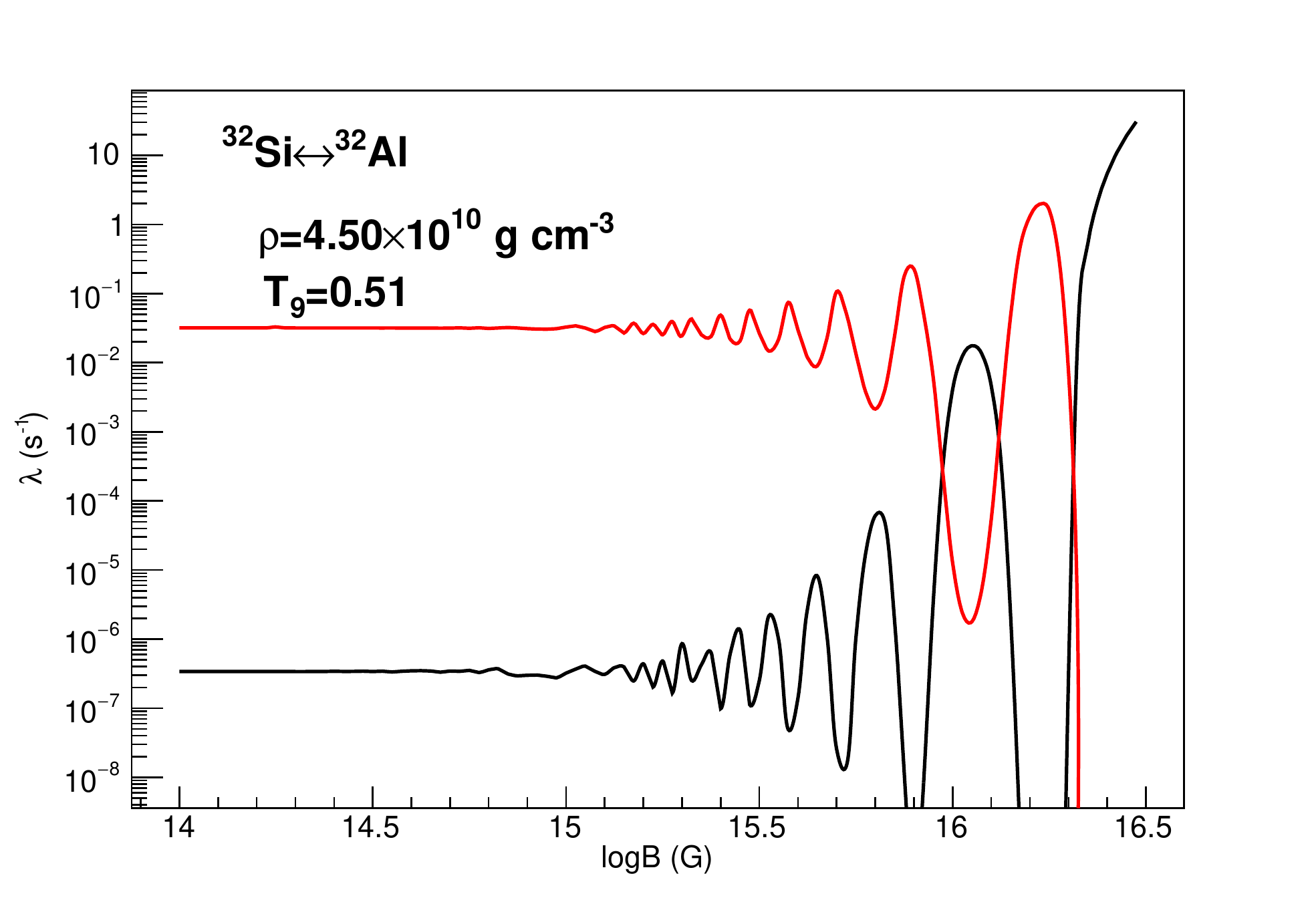}{0.5\textwidth}{32Si}
    }
	\caption{Evolution of Urca pair formation for $^{29}$Mg$\leftrightarrow^{29}$Na (left)
		and $^{32}$Si$\leftrightarrow^{32}$Al (right) at the indicated temperatures and densities.
	}
    \label{mg29_fig}
\end{figure}

 The introduction of magnetic fields can also create Urca pairs where they would not
have otherwise existed.  This is shown in Figure \ref{mg29_fig}, for two pairs at densities
and temperatures not conducive to the formation of Urca pairs for the nuclei shown at low
fields.  However, as the field increases, the rates can match and possibly form Urca pairs.  The
additional  degree of freedom from the  magnetic field extends the range of possible conditions at
which Urca pairs can form within the NS crust.  

\begin{figure}
    \centering
    \gridline{
    \fig{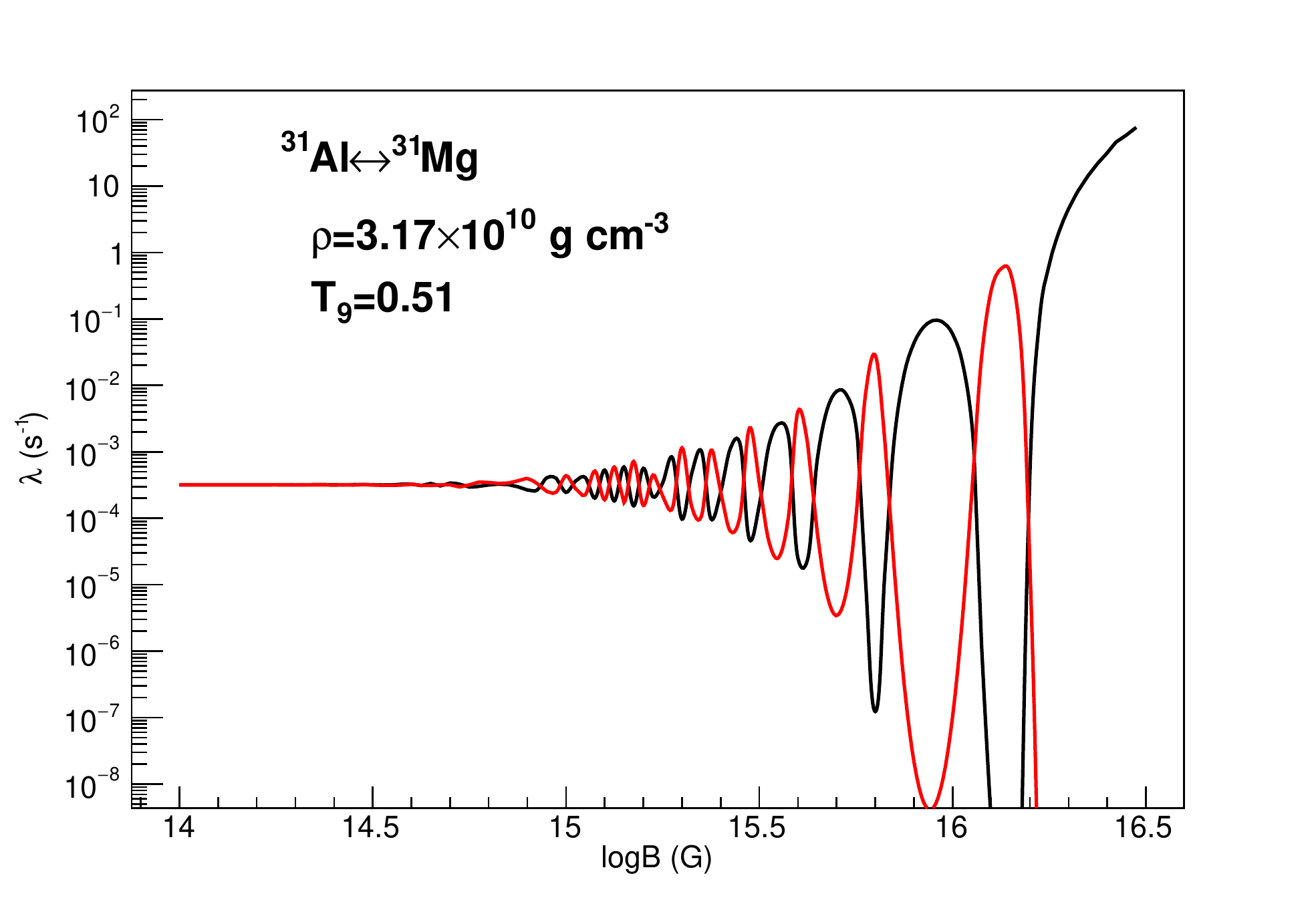}{0.5\textwidth}{GS}
    }
     \gridline{
        \fig{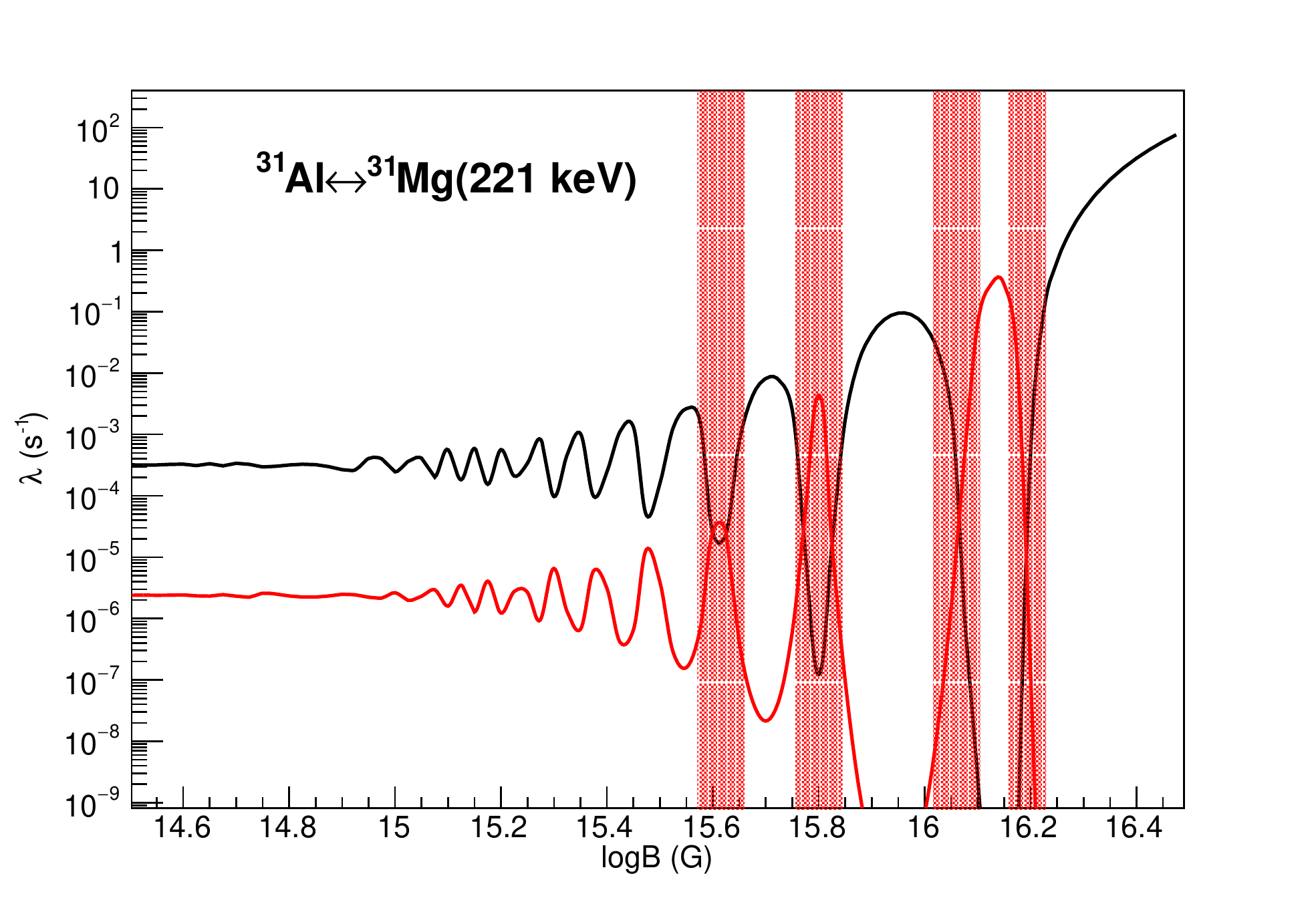}{0.5\textwidth}{1st Excited State}
    \fig{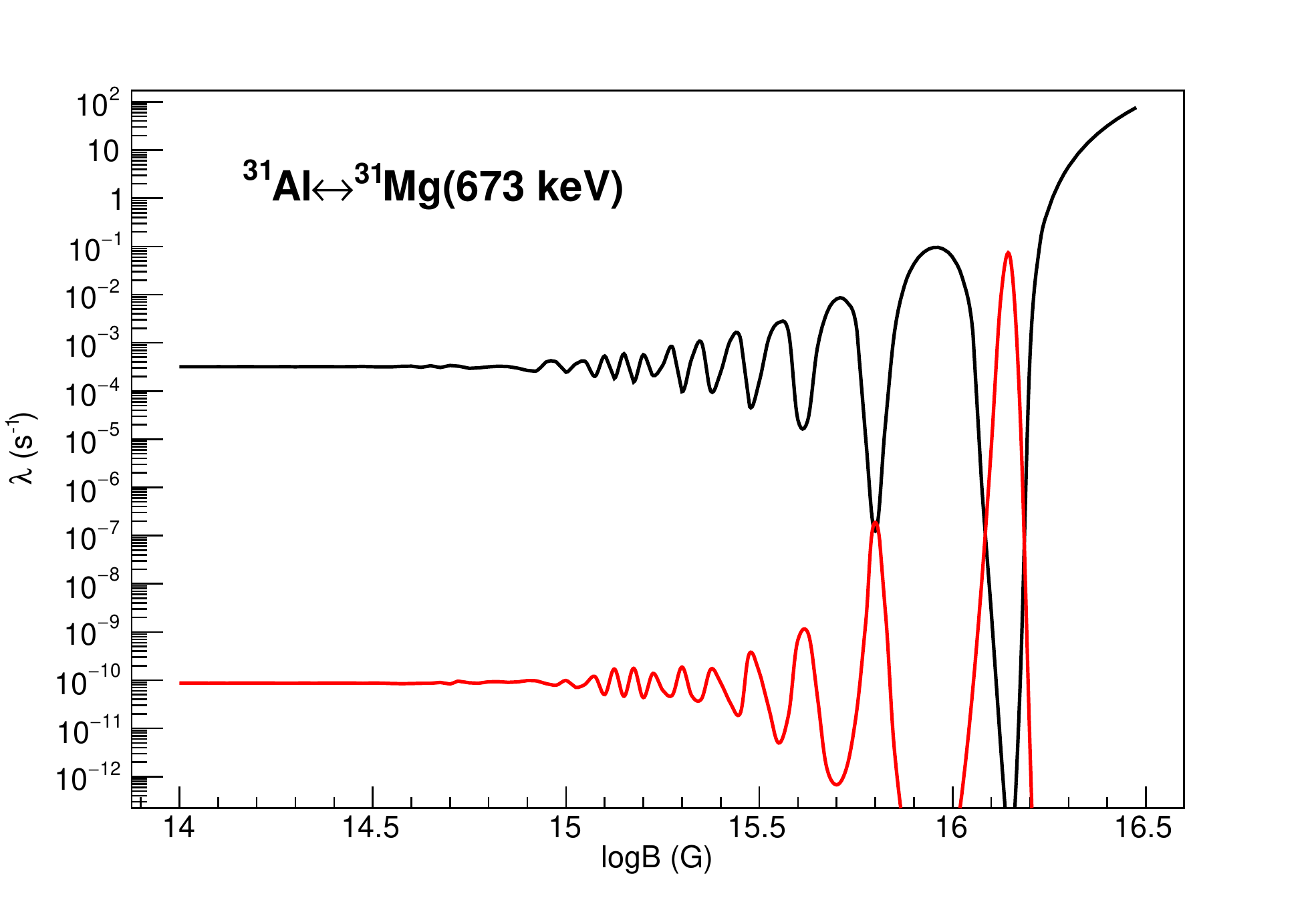}{0.5\textwidth}{2nd Excited State}
    }
	\caption{EC and $\beta^-$ rates for the $^{31}$Al$\leftrightarrow^{31}$Mg Urca pair
	for transitions to the ground state in $^{31}$Mg (top), the first excited state (bottom left),
 	and the second excited state (bottom right).  The red bands in the bottom left panel
 	indicate fields where Urca pairing may be more efficient.}
    \label{Mg31_fig}
\end{figure}

Finally,  in addition to beta decay, Urca pairs can form for EC daughter excited states.  For excited states in
the EC daughter nucleus, the EC rate drops.  The daughter nucleus de-excites almost immediately
and decays via $\beta^-$ decay back to the original nucleus.  However, because the
$\beta^-$ rate exceeds the EC rate for excited states, the cooling efficiency is lower.
The rate oscillations, however, may open Urca pair transitions to excited states in the
EC daughter nucleus.   This is shown in Figure \ref{Mg31_fig}.  EC and $\beta^-$ rates
to the ground state of $^{31}$Mg  are shown as a function of magnetic field, as well as 
to the first two excited states of $^{31}$Mg.  While Urca pairs including excited states
may not be possible at  low field strength, higher fields may allow for Urca pairs between
excited states.   This is particularly interesting because EC to excited states may
open up transition rates that are more favored.

\begin{figure}
    \centering
   \gridline{
   \fig{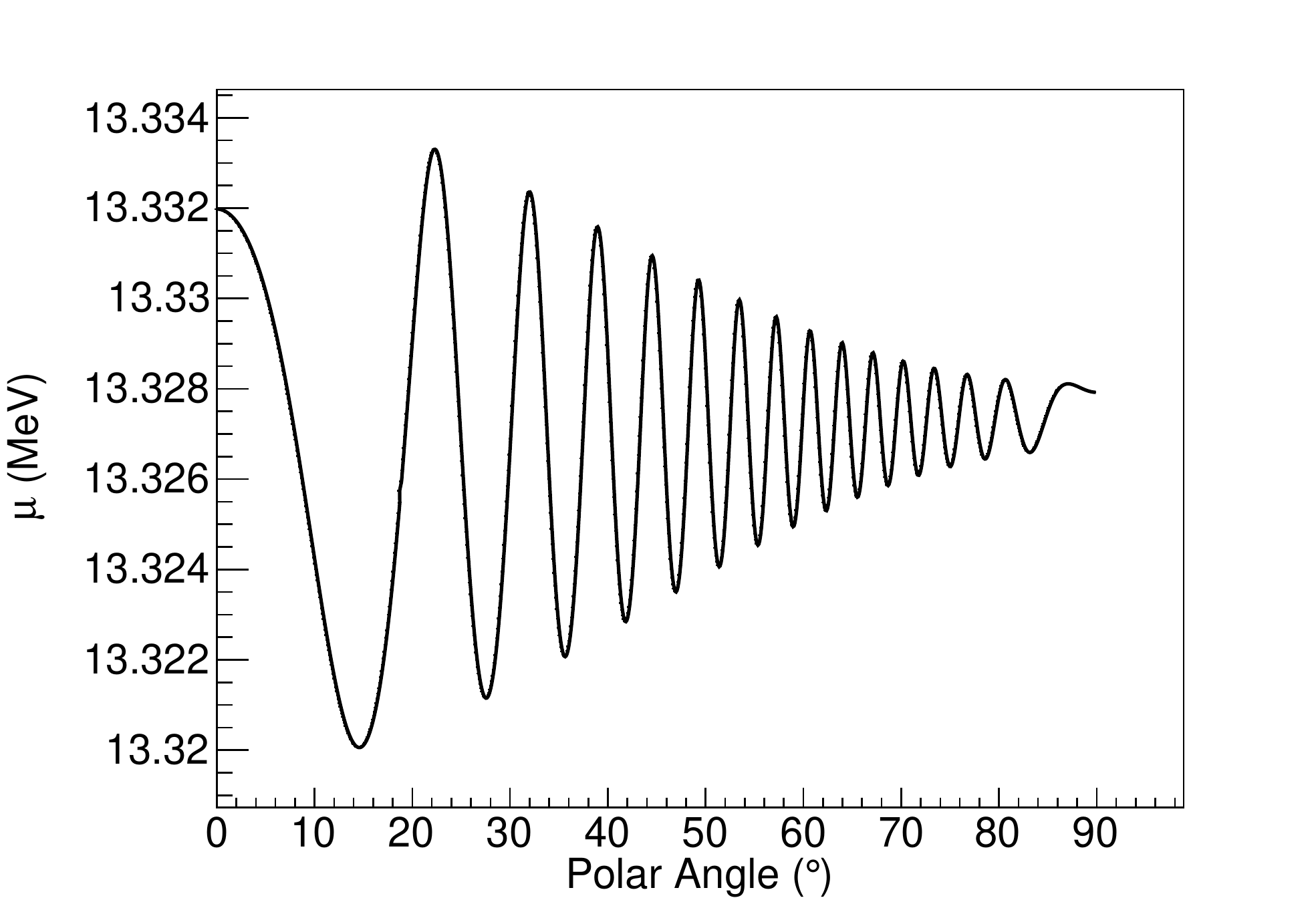}{0.33\textwidth}{}
     \fig{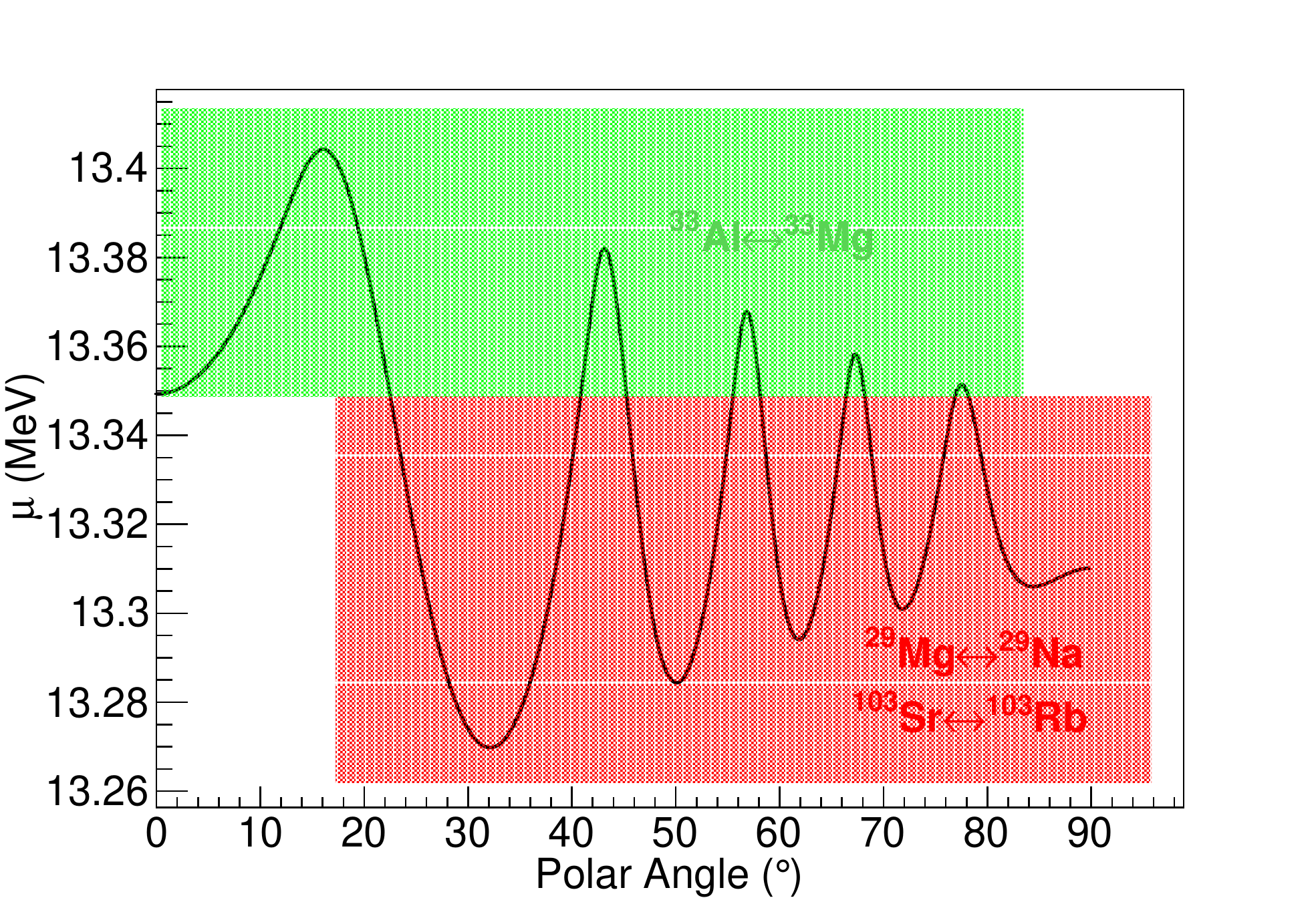}{0.33\textwidth}{}
       \fig{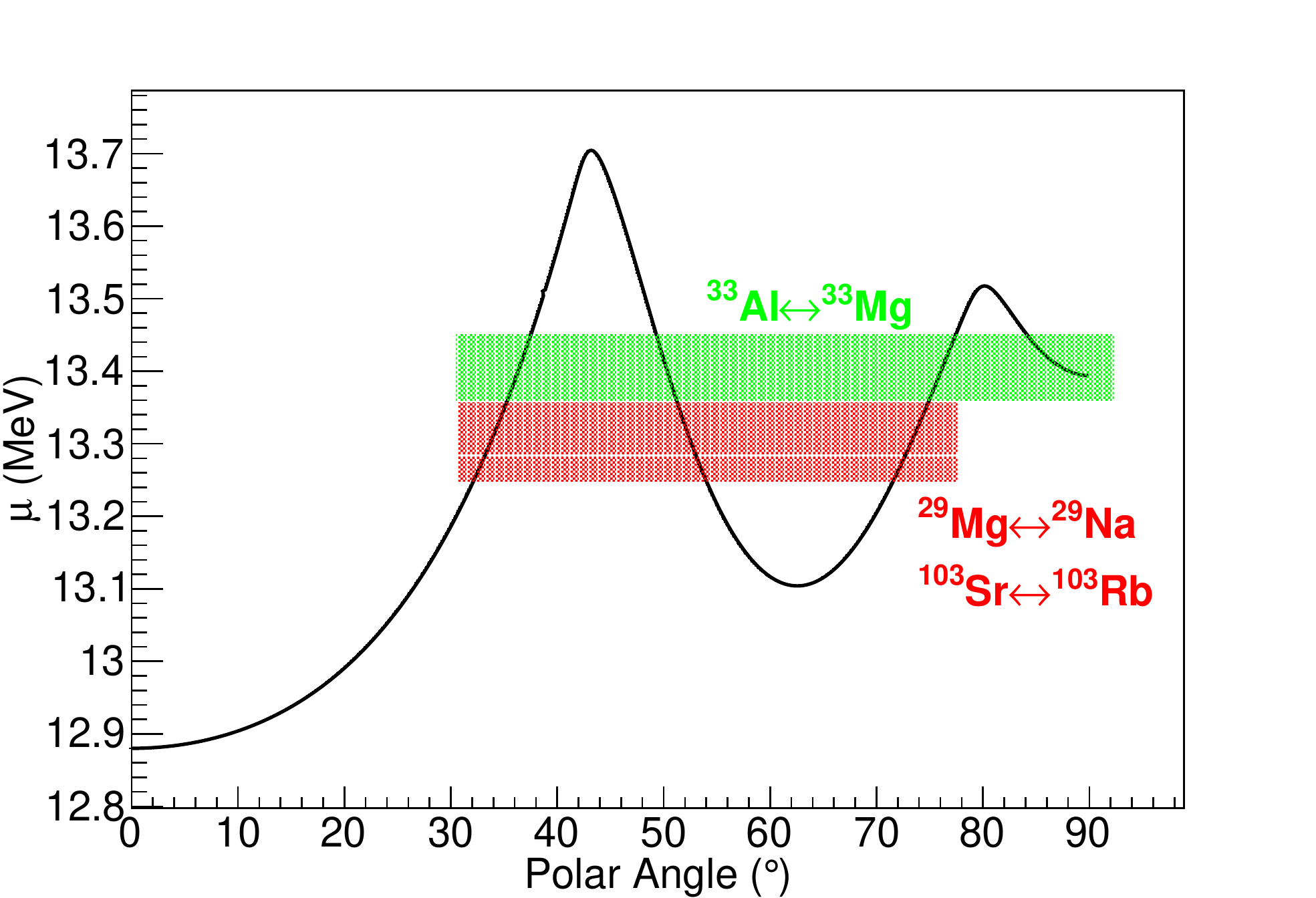}{0.33\textwidth}{}
   }
    \gridline{
        \fig{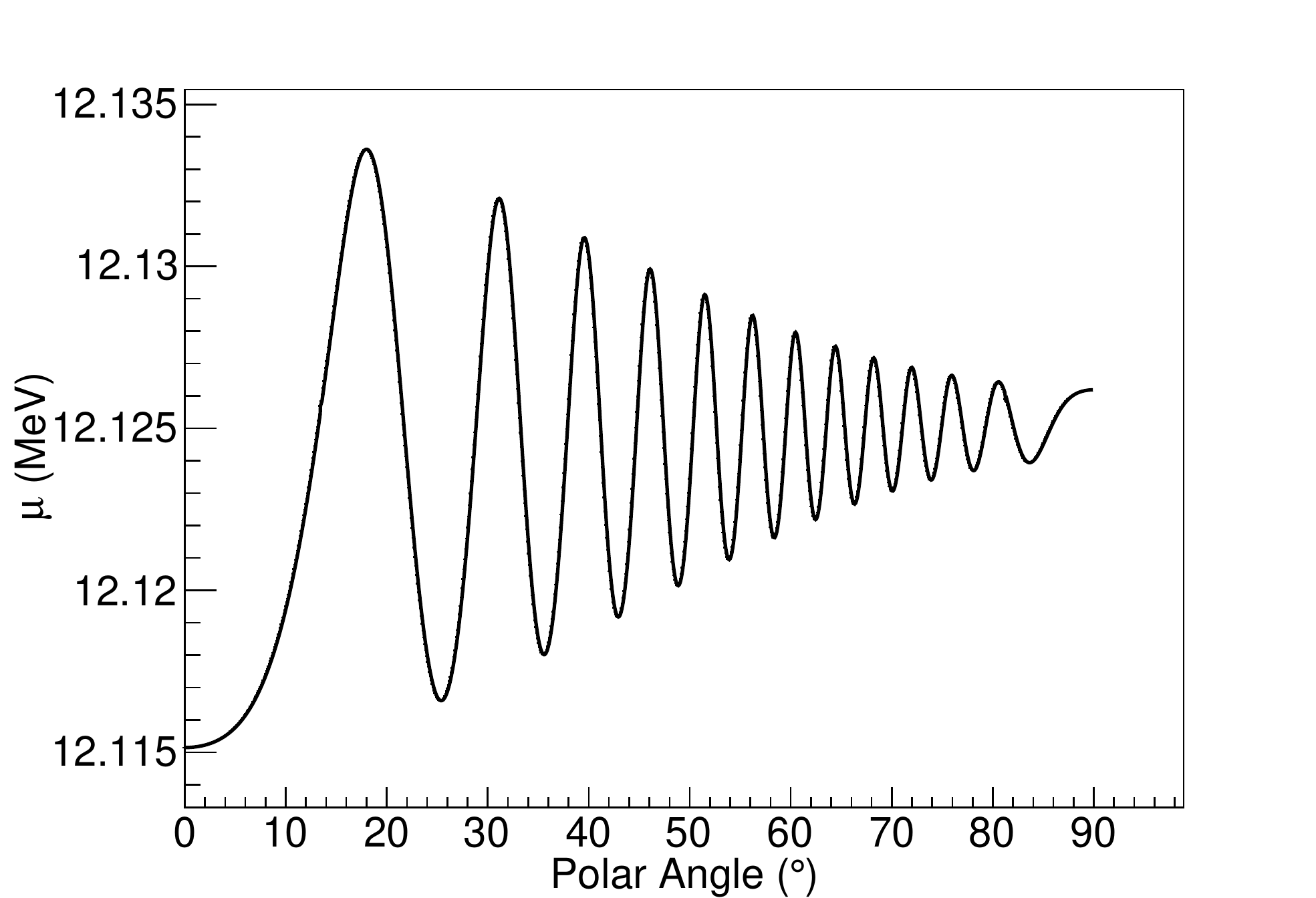}{0.33\textwidth}{}
        \fig{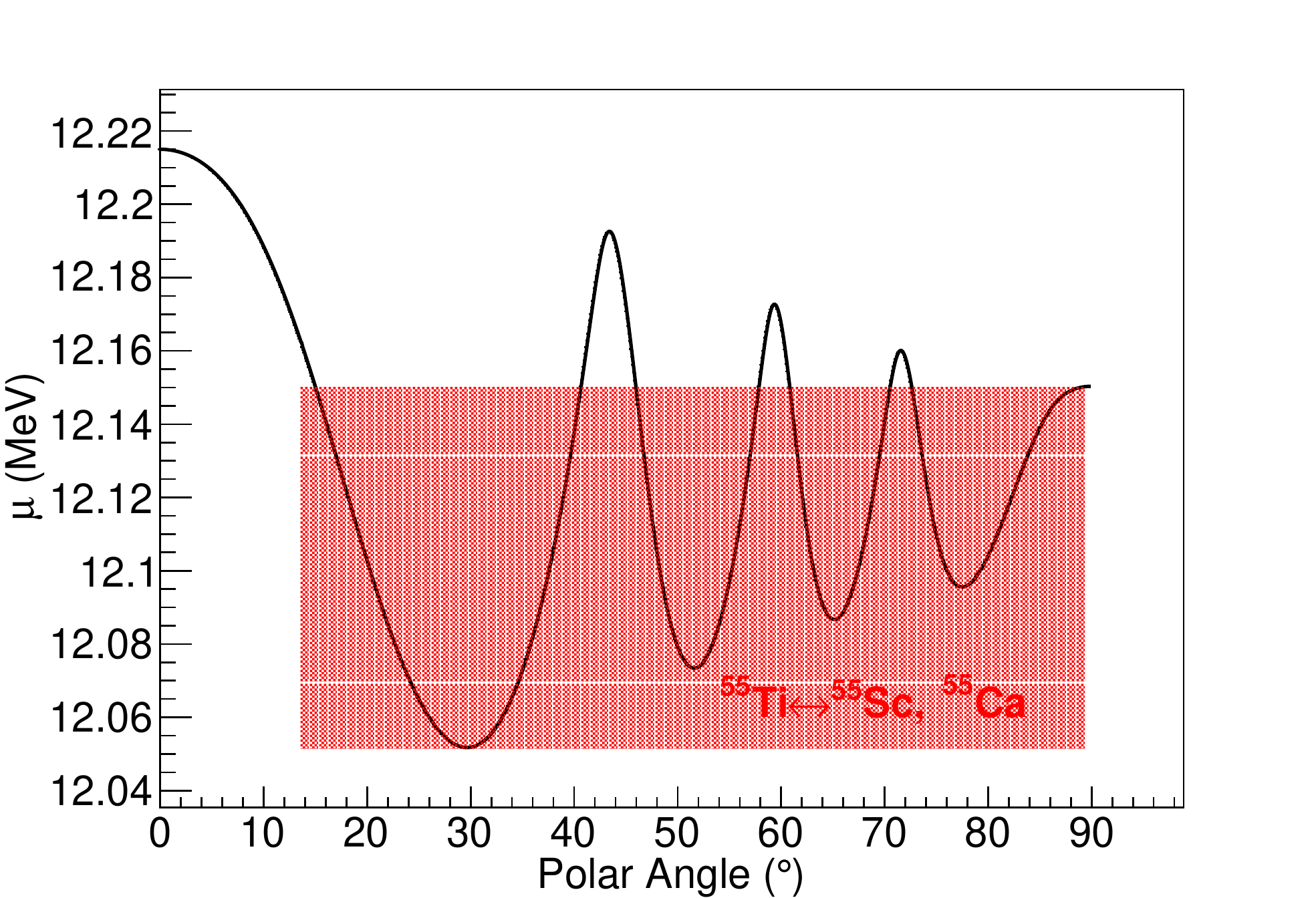}{0.33\textwidth}{}
       \fig{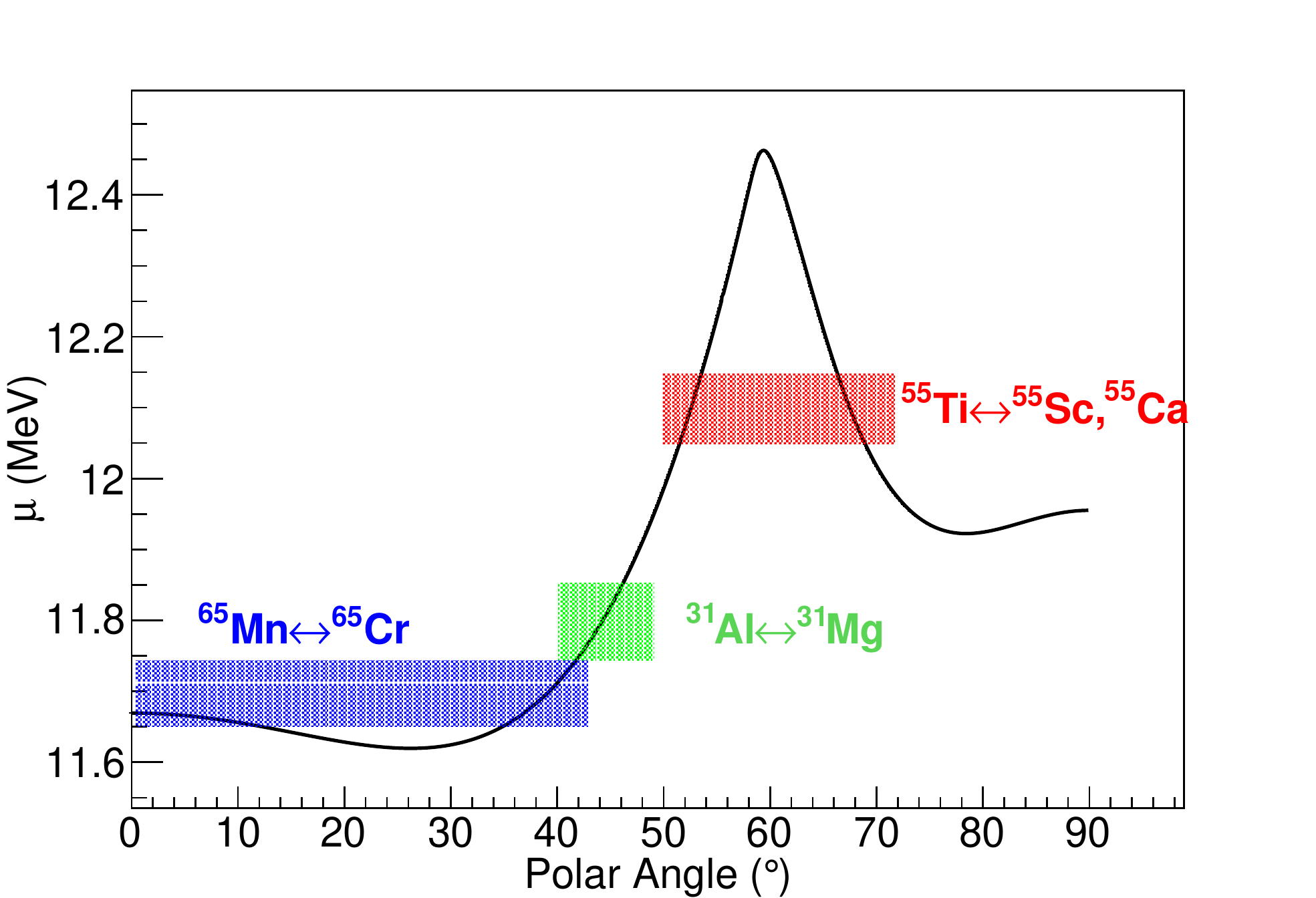}{0.33\textwidth}{}
   }
    \caption{
    Prevalence of Urca pairs for two different values of $\rho Y_e$ at various magnetic
    fields.  Top row: $\rho Y_e=1.7\times 10^{10}$ g cm$^{-3}$, Bottom row: $\rho Y_e=1.3\times 10^{10}$ g cm$^{-3}$.  Values are shown for magnetic fields of 10$^{15}$ G (left), 10$^{15.5}$ G (center), and 
    10$^{16}$ G. Shaded regions are regions where the indicated Urca pair dominates.  For the 
    left column, the dominant Urca pair is $^{29}$Mg$\leftrightarrow^{29}$Na and $^{103}$Sr$\leftrightarrow^{103}$Rb (top) and 
    $^{55}$Ti$\leftrightarrow^{55}$Ca,$^{55}$Ca (bottom).
    }
    \label{mu_theta_fig}
\end{figure}

As an example, Figure \ref{mu_theta_fig} shows the electron
chemical potential as a function of polar angle for a layer of constant density and temperature within 
the crust.  For all panels in this figure, a  typical temperature of $T_9=0.51$ is adopted.  For the top row
of panels, $\rho Y_e=4.70\times 10^{10}$ g cm$^{-3}$, and for the bottom row $\rho Y_e=3.73\times 10^{10}$
g cm$^{-3}$.

The shaded regions in these figures indicate those for which the chemical potential is within $kT$ of 
the EC Q-values of the indicated reactions, $Q_{EC}-kT\le\mu_e\le Q_{EC}+kT$.  With this evaluation, 
an Urca pair is not consistent across the entire surface of the NS at a given temperature and density.  At high 
enough magnetic fields, an Urca pair may dominate a particular angular band.  This may also have
the effect of making additional Urca pairs possible in various angular regions.

\begin{figure}
    \centering
    \gridline{
        \fig{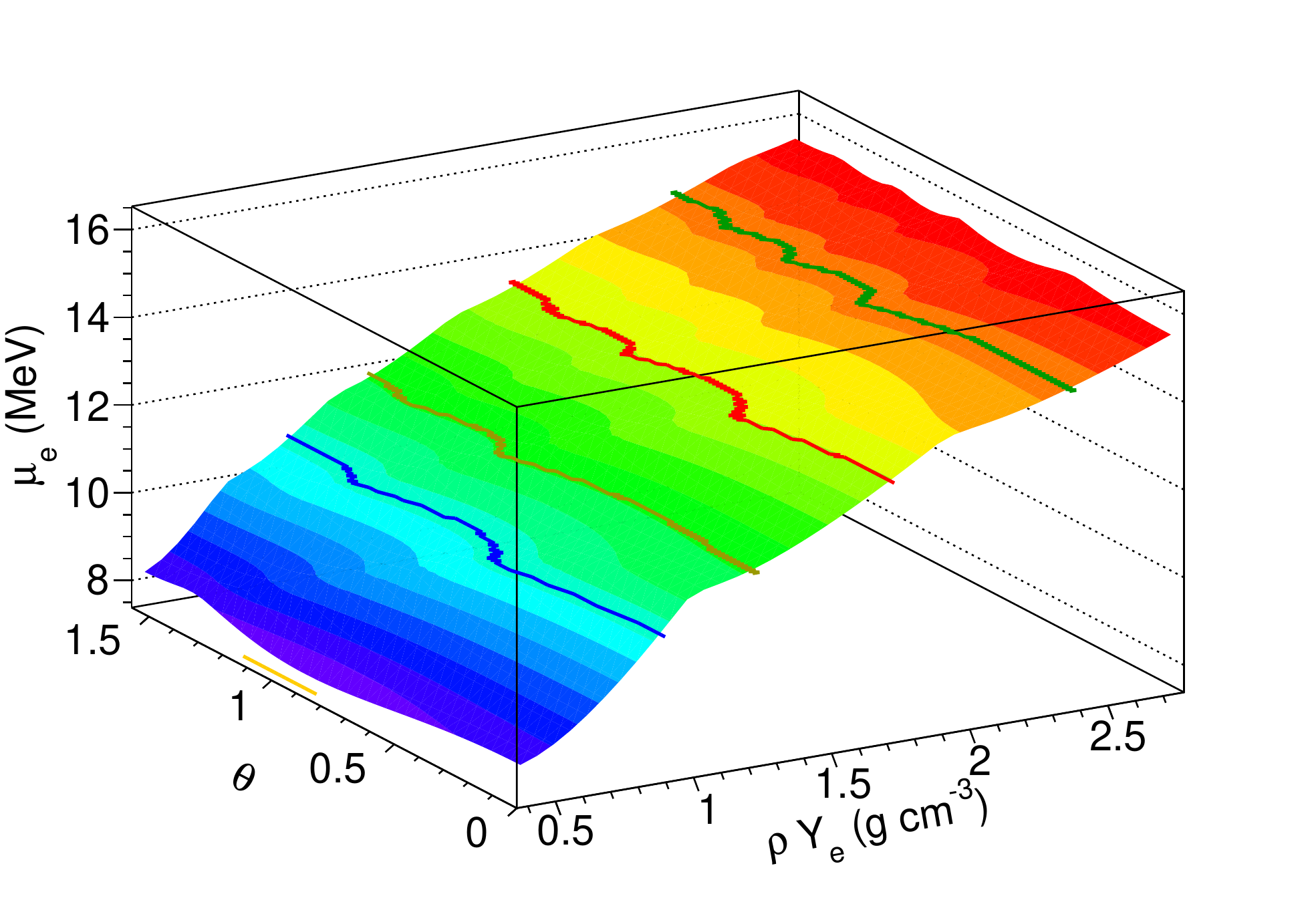}{0.5\textwidth}{}
        \fig{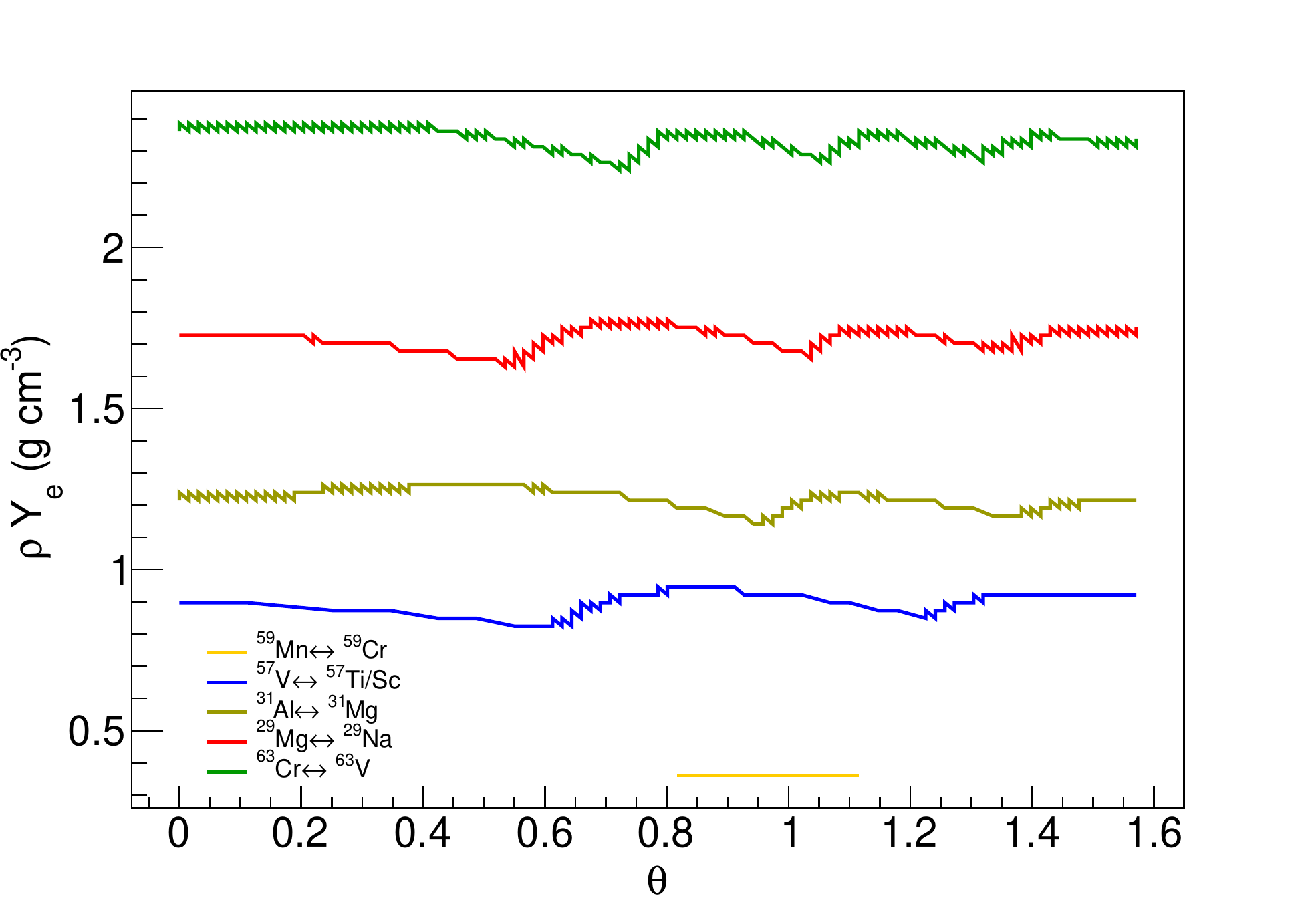}{0.5\textwidth}{}
    }
    \gridline{
        \fig{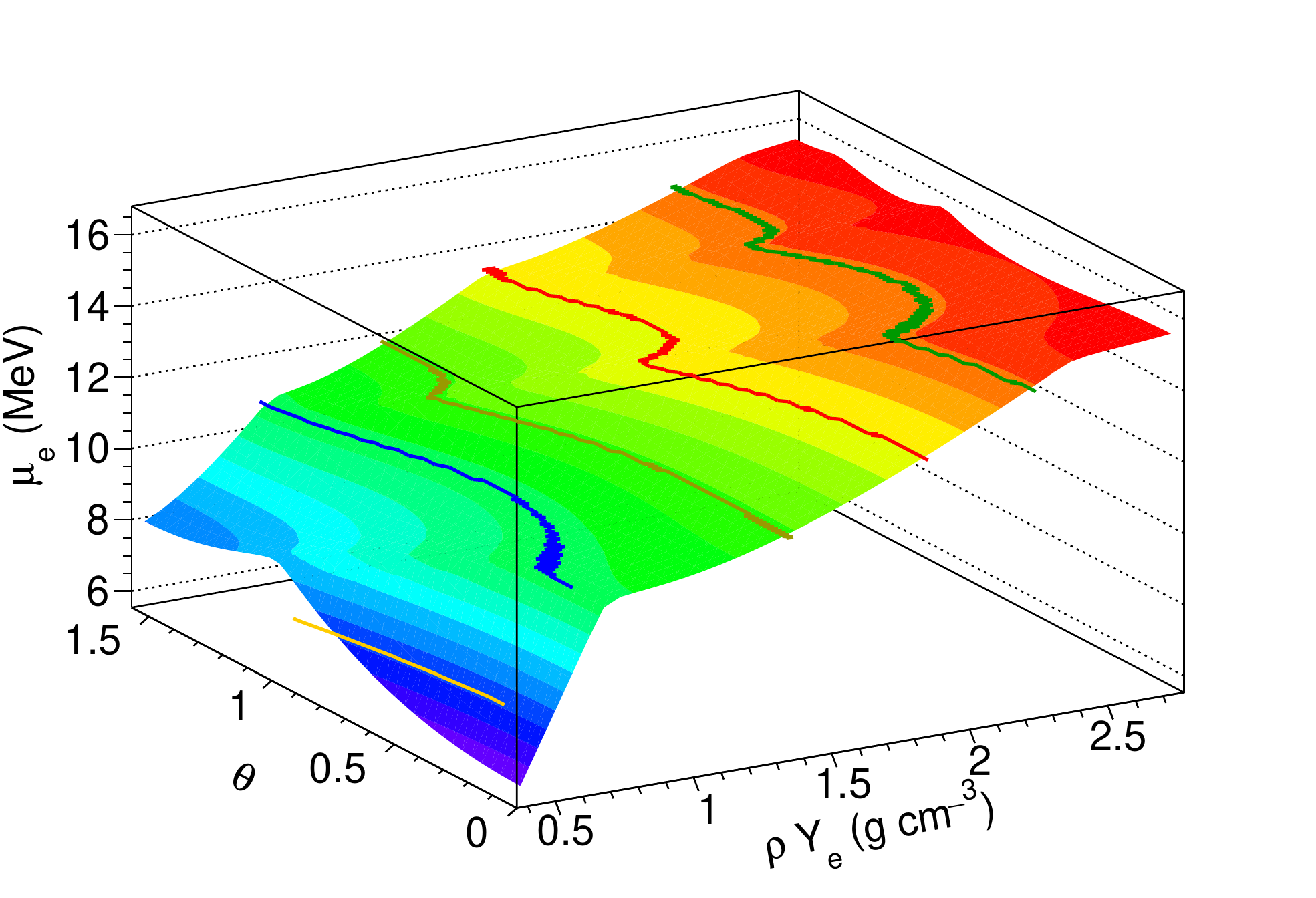}{0.5\textwidth}{}
        \fig{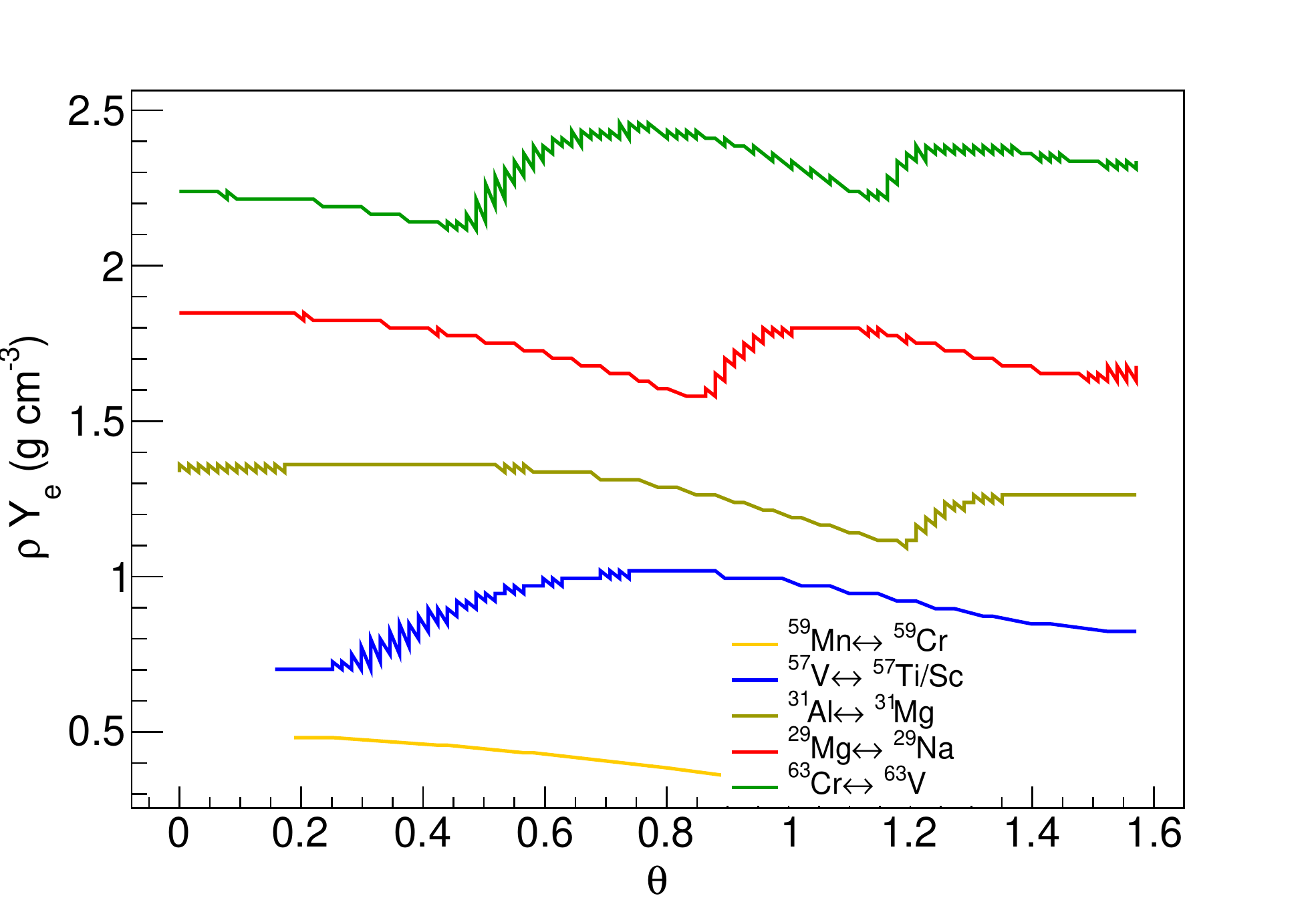}{0.5\textwidth}{}
    }
    \gridline{
        \fig{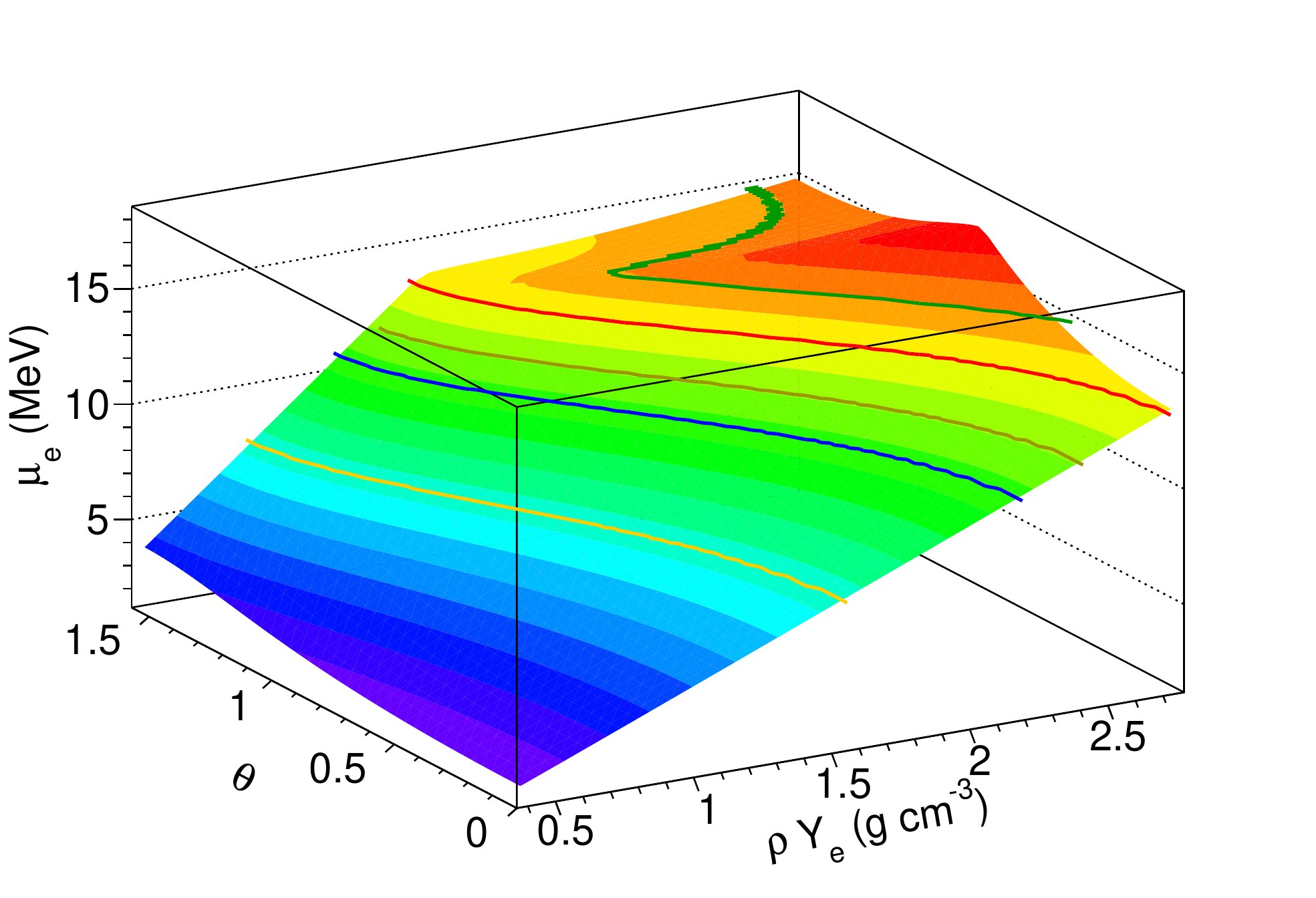}{0.5\textwidth}{}
        \fig{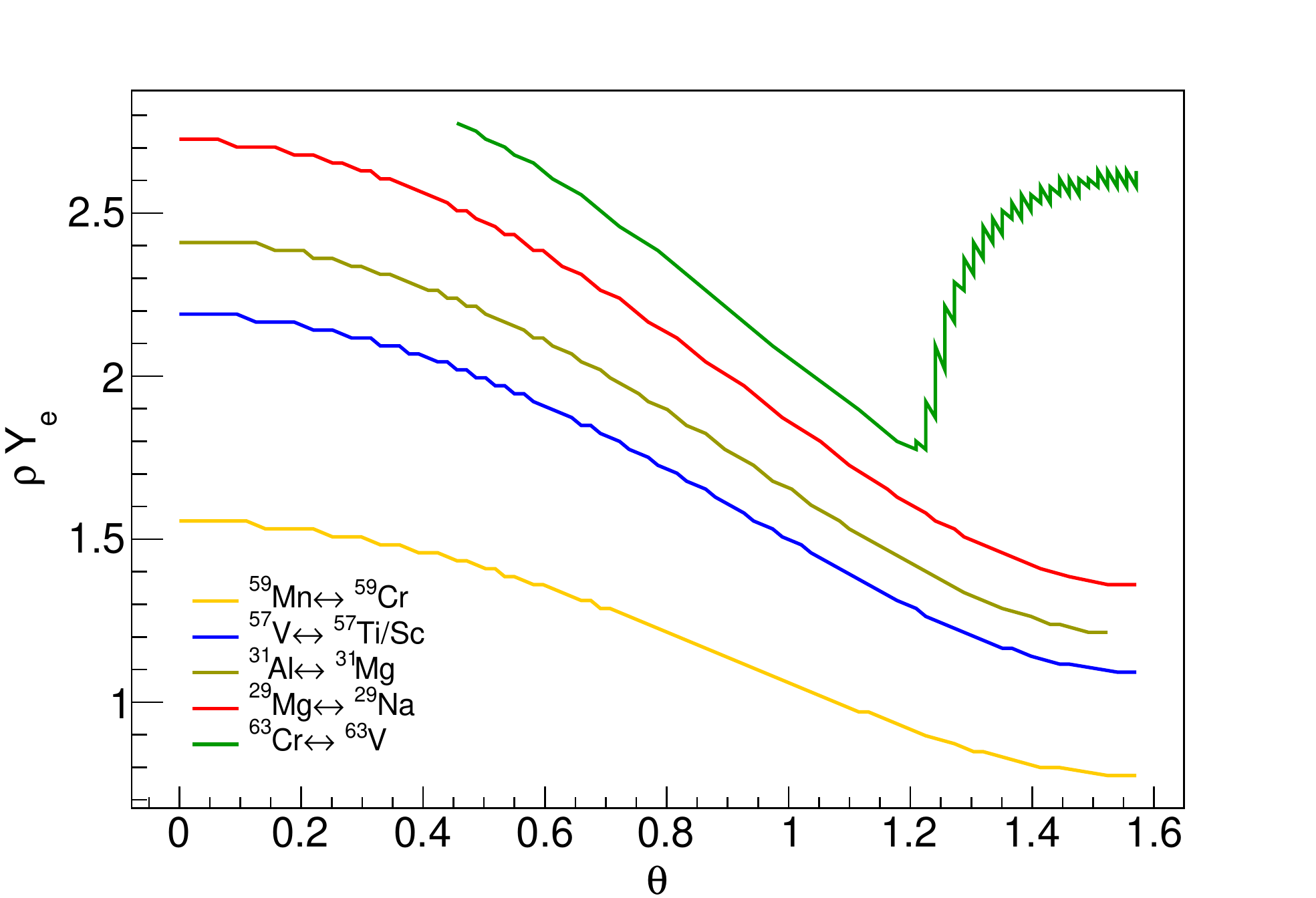}{0.5\textwidth}{}
    }
    \caption{
    Left: Electron chemical potential as a function of $\rho Y_e$ and polar
    angle for a neutron star with a dipole field of 10$^{15.75}$ G (top), 10$^{16}$ G (middle), and 10$^{16.5}$ G bottom.  Units of $\rho Y_e$ are 10$^{10}$ g cm$^{-3}$.  Right:  The electron density $\rho Y_e$ and polar angle for the same magnetic fields at which various crustal Urca pairs are  present.  These correspond to the lines in the left column.}
    \label{mu_angle_dens}
\end{figure}

Because of the shift in electron chemical potential with magnetic field, the
location or Urca pairs within the neutron star crust varies with field and 
latitude (polar angle on the stellar surface). As an example, consider Figure
\ref{mu_angle_dens}, which shows the electron chemical potential as a function of
polar angle for various surface magnetic fields (at the poles), latitudes, and 
densities within a NS crust.  Because the chemical potential changes with 
both field and density, the physical location of Urca pairs may change.  
This will also result in a differentiation in the neutrino luminosity 
of various stars.  For example, for a very large field of 10$^{16.5}$ G as indicated
in the figure, the density for which an Urca pair occurs is highest at the poles
of the star, resulting in a larger polar neutrino luminosity, whereas, the 
stellar equator will form Urca pairs at a lower density.  However, for
 the $^{63}$Cr$\leftrightarrow ^{63}$V Urca pair, it can be seen that the minimum luminosity 
occurs at $\theta\approx$ 1.2 rad.  The magnetic field can result in a shift in the locations of Urca pairs as a function of latitude as seen in this figure.  It 
can also be seen that magnetic fields will enable the existence of new pairs.

For a constant density and temperature, changes in magnetic field can result
in a change in the electron chemical potential (Equation \ref{derv}).  This changes
the overall phase space for electrons in $\beta^-$ decay and electron capture.
If an Urca pair is defined as a $\beta^-$-EC pair for which $\mu_e\sim Q$, then the 
density at which an Urca pair can exist also depends on the environmental
magnetic field.

\begin{figure}
    \gridline{
        \fig{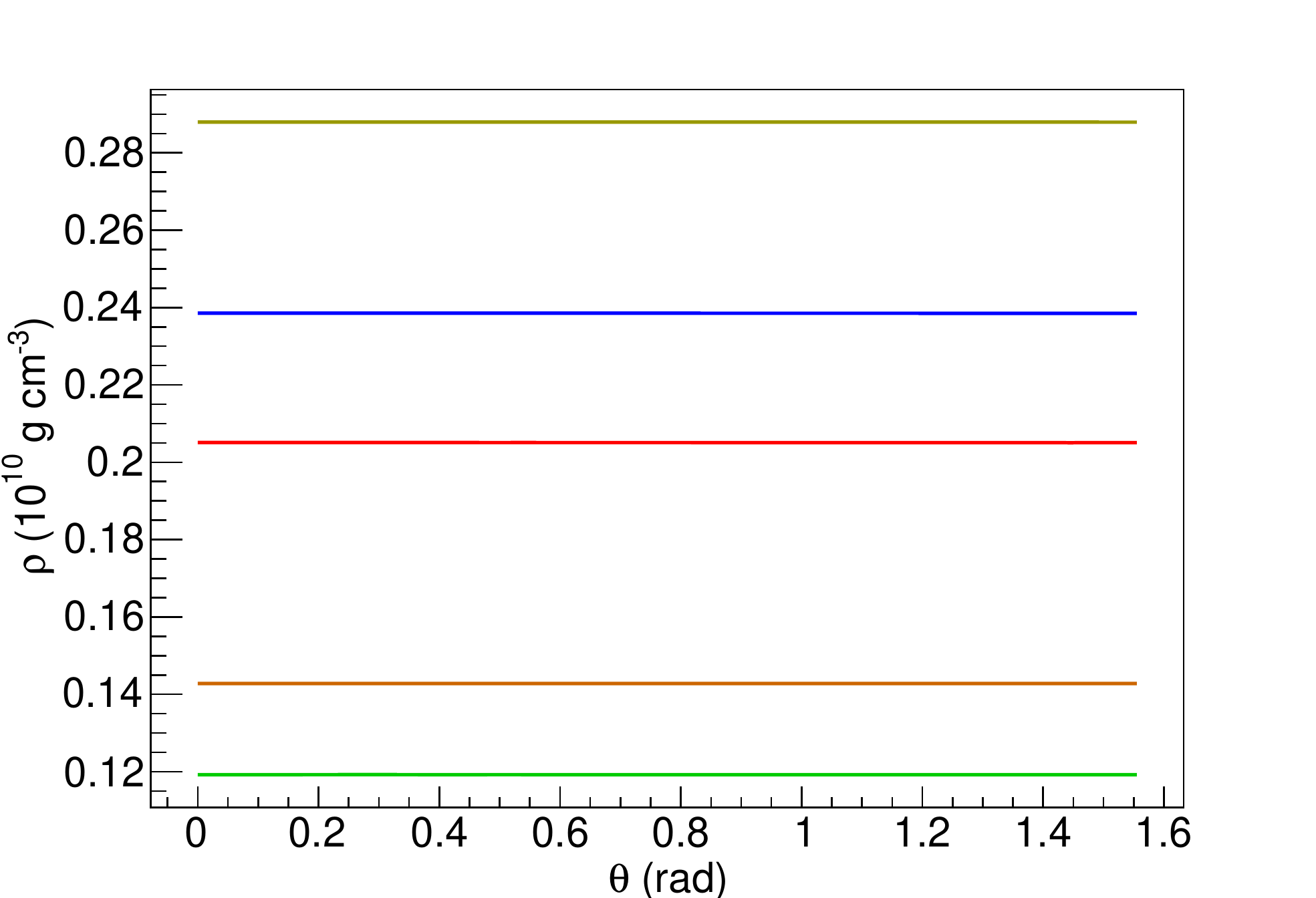}{0.33\textwidth}{}
        \fig{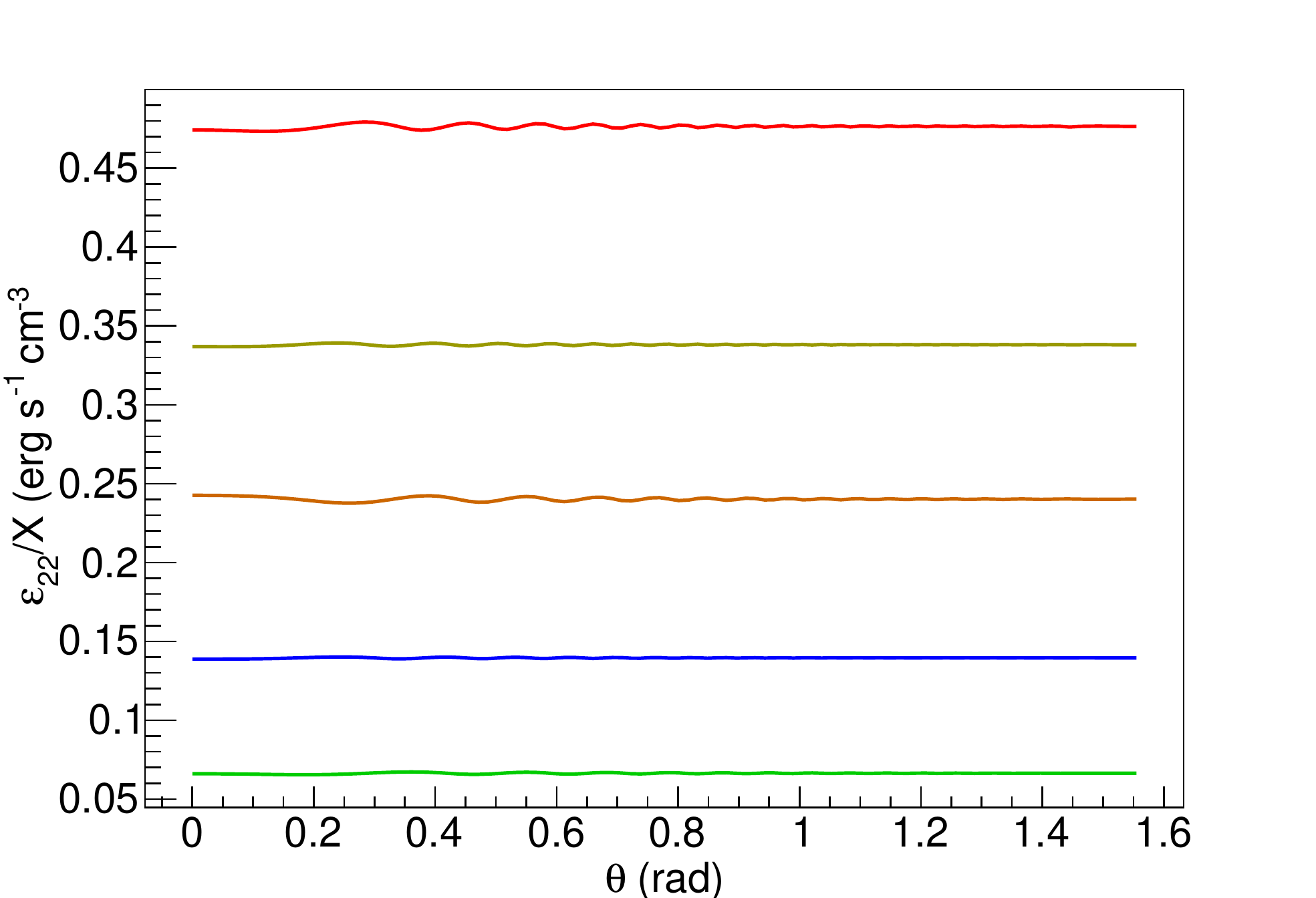}{0.33\textwidth}{}
        \fig{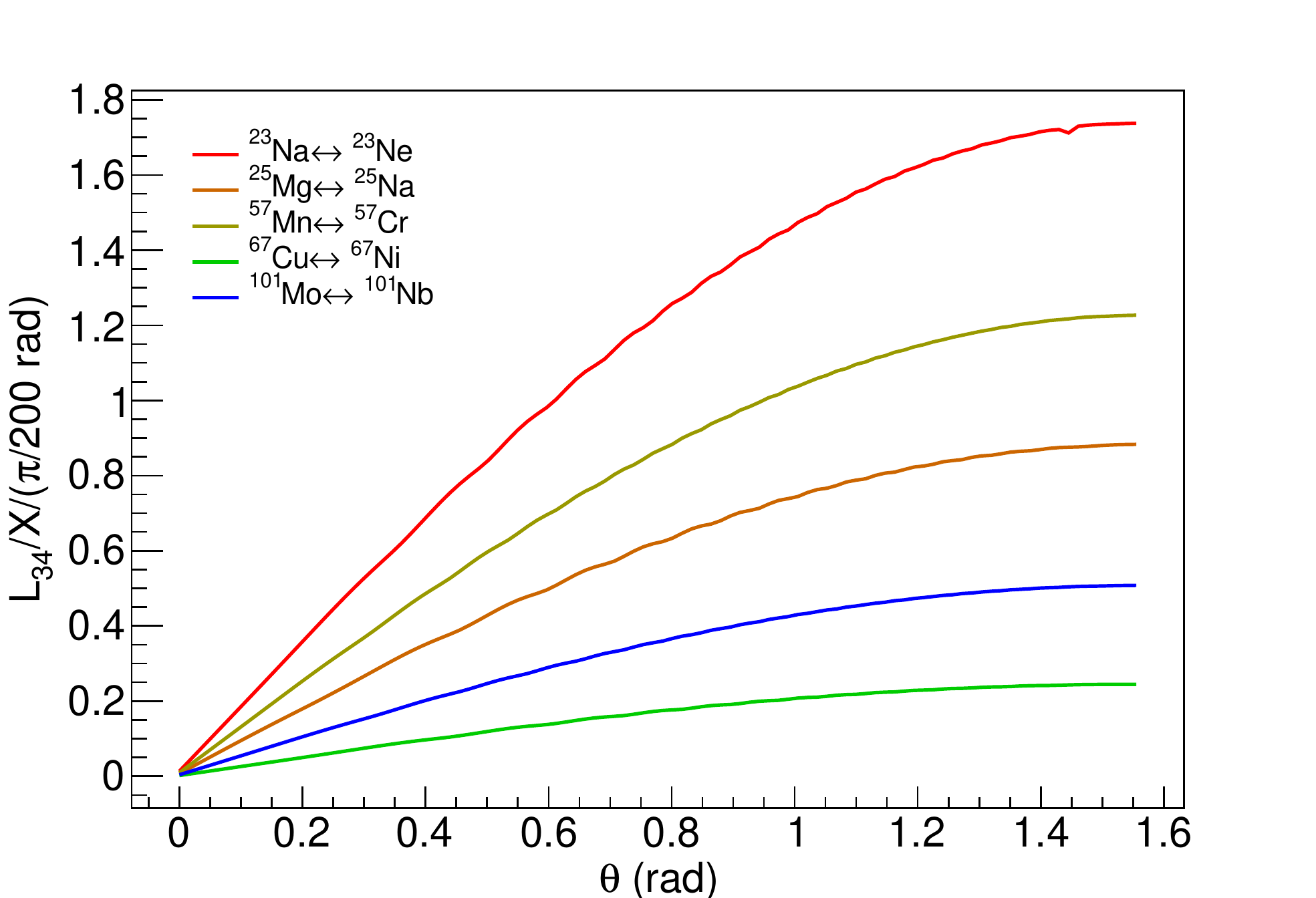}{0.33\textwidth}{}
    }
    \gridline{
        \fig{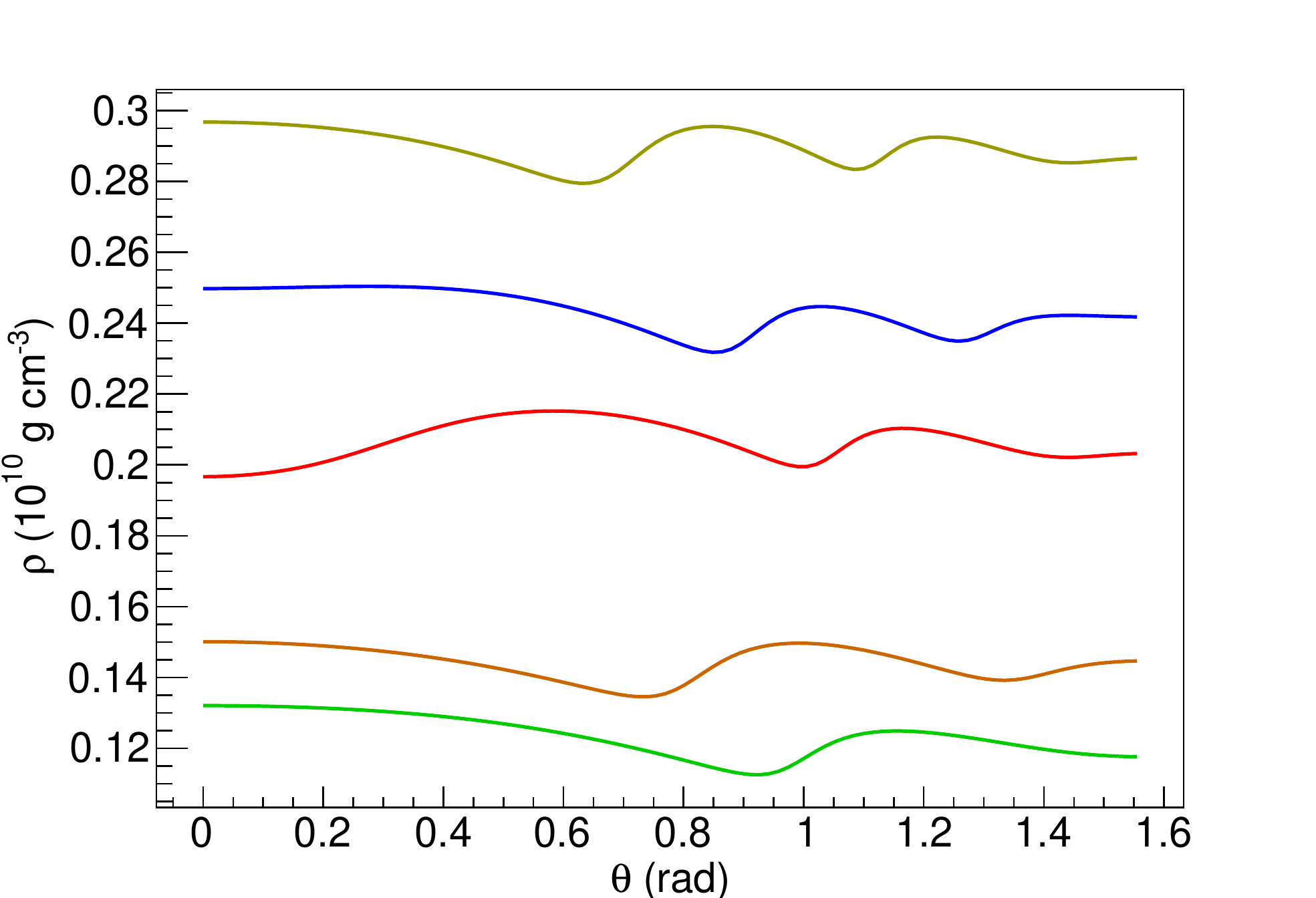}{0.33\textwidth}{}
        \fig{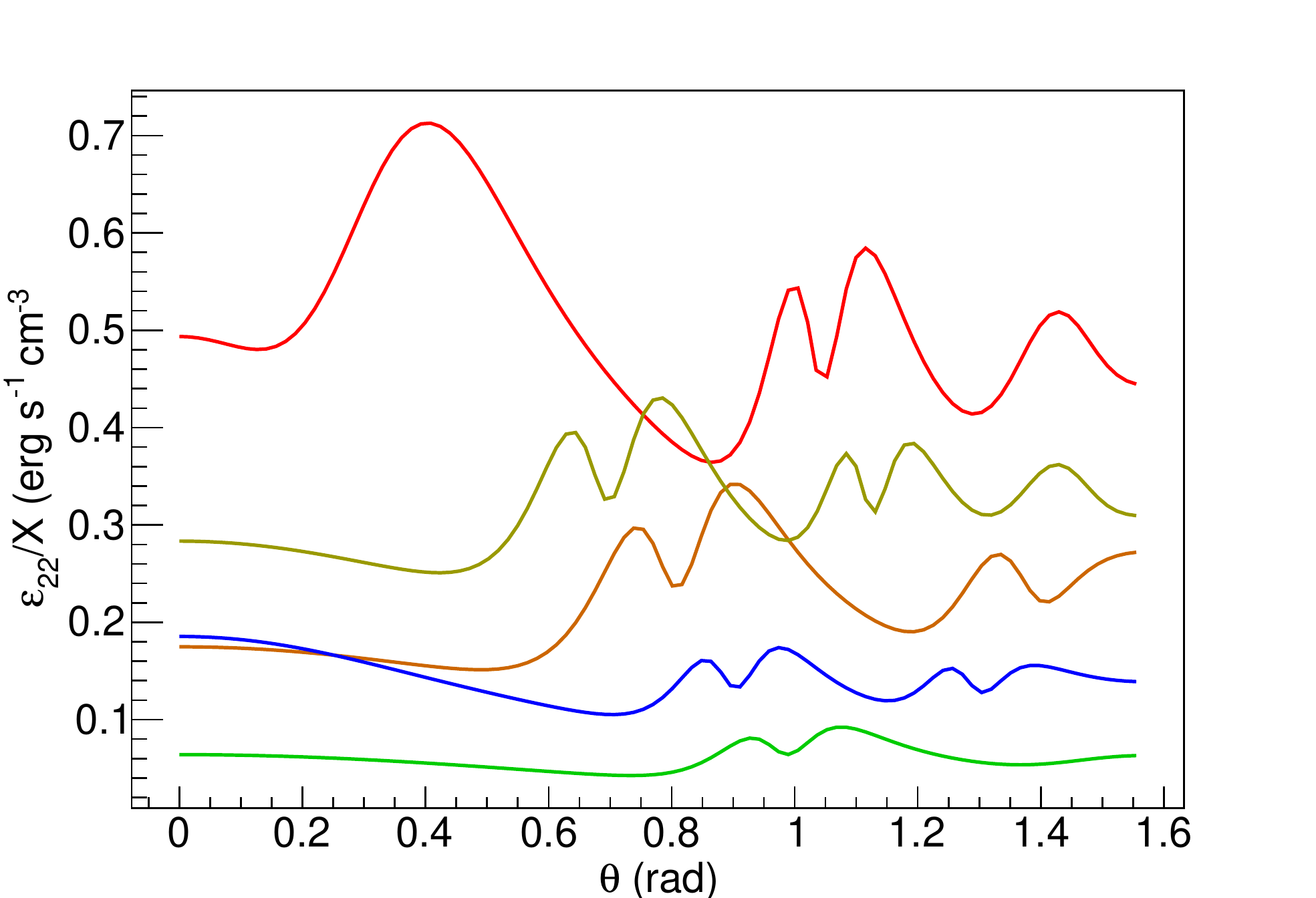}{0.33\textwidth}{}
        \fig{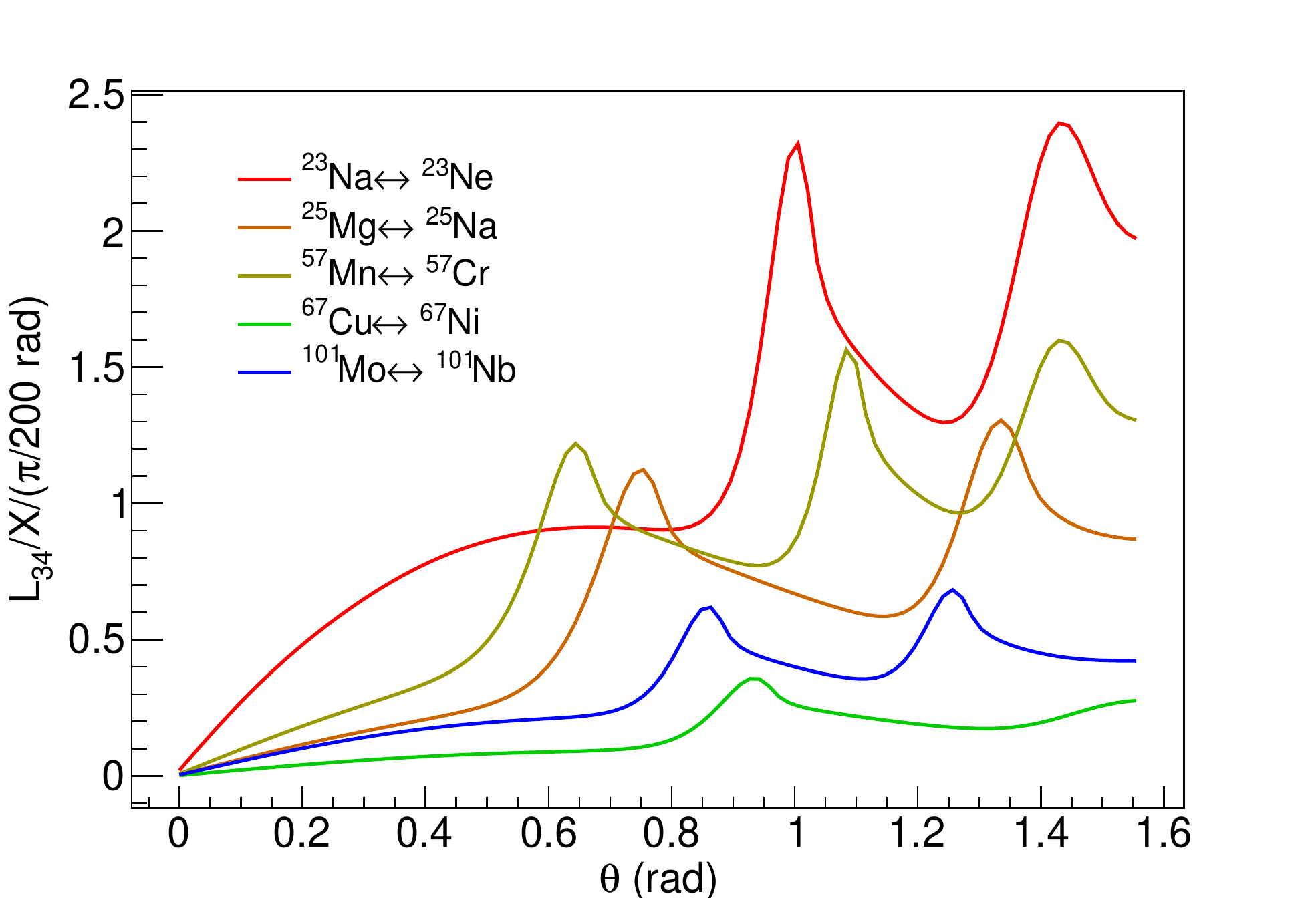}{0.33\textwidth}{}
    }
     \gridline{
        \fig{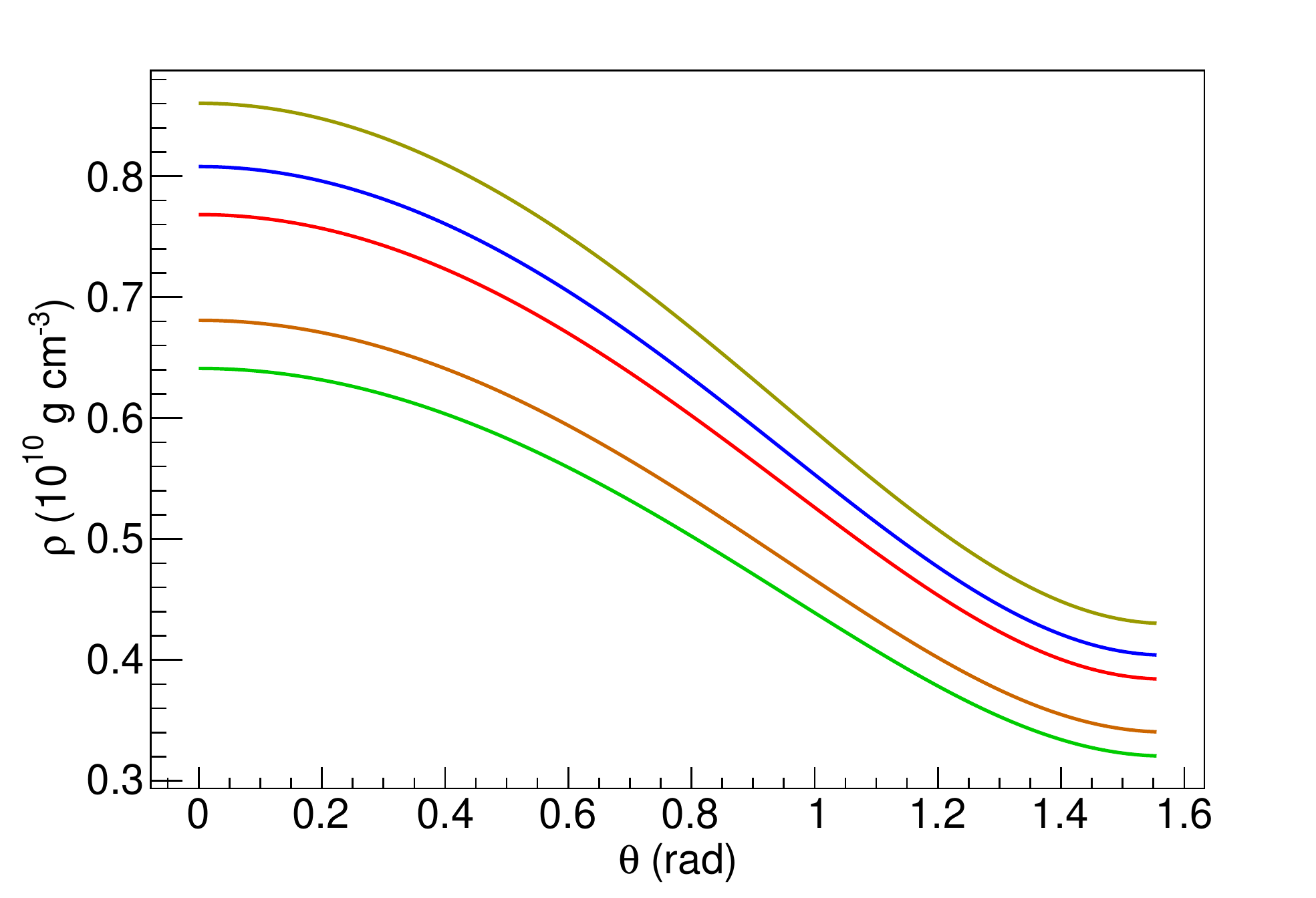}{0.33\textwidth}{}
        \fig{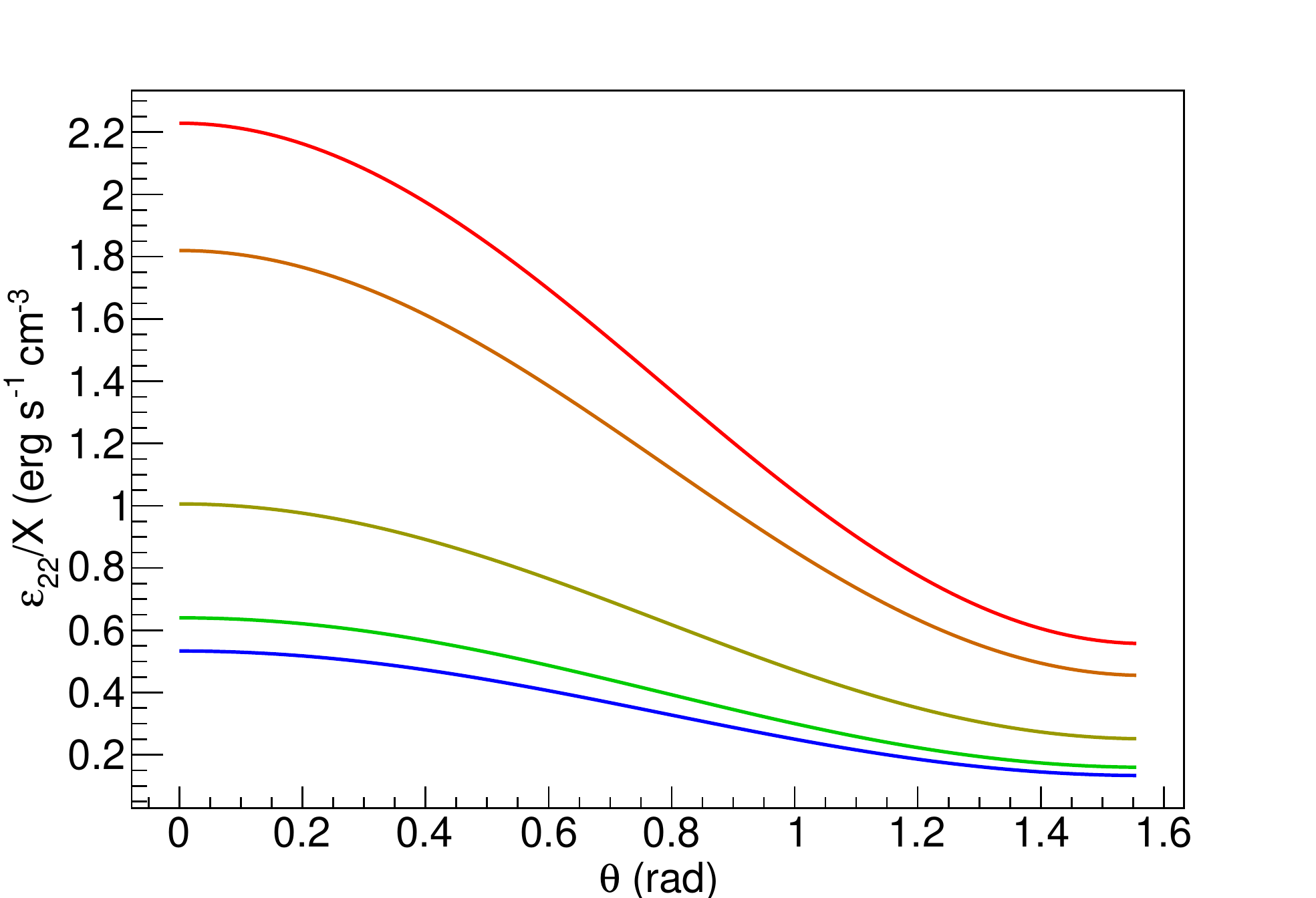}{0.33\textwidth}{}
        \fig{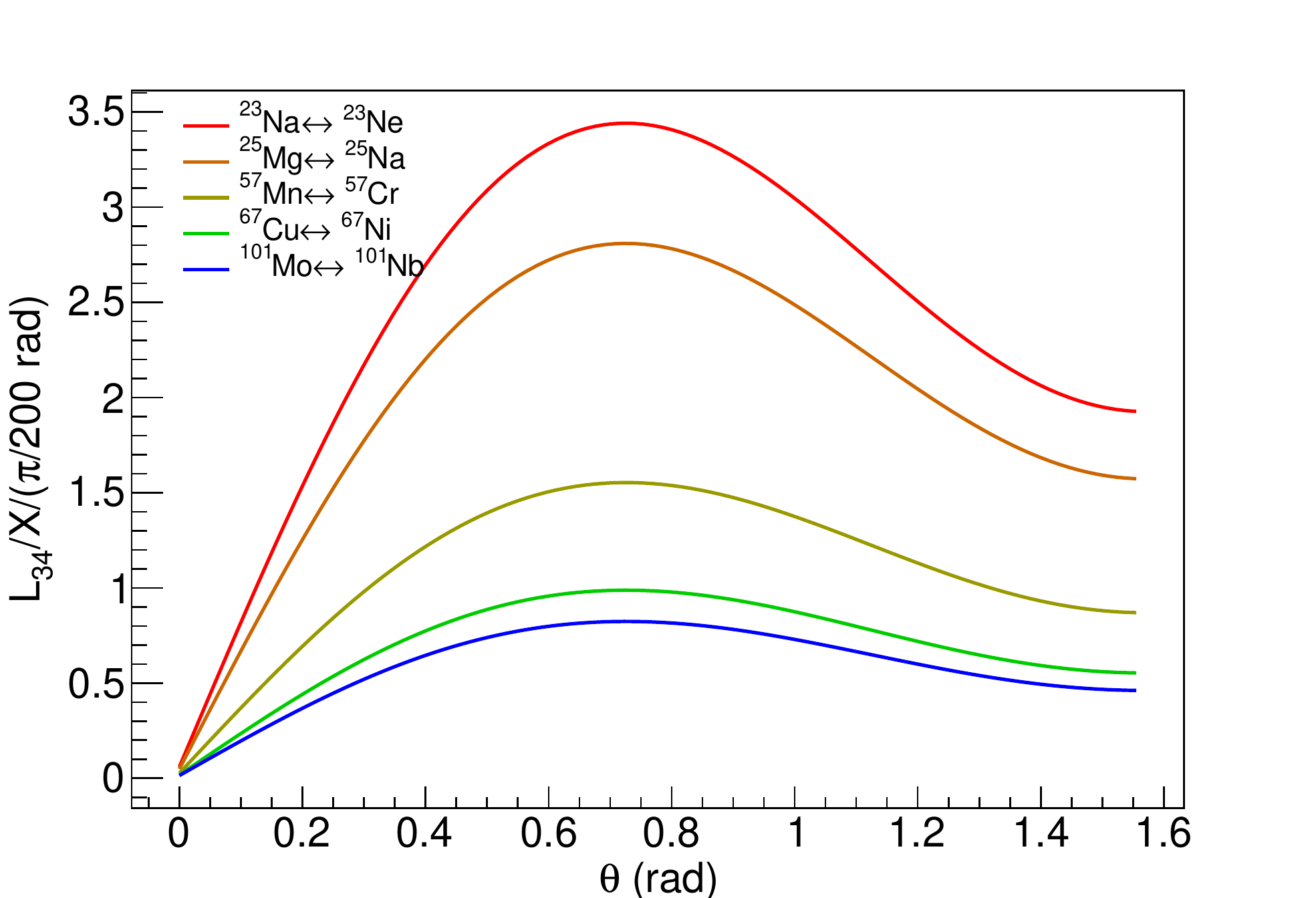}{0.33\textwidth}{}
    }
    \caption{Left  panels: The density at which an Urca pair is viable for various ocean Urca pairs.  Center  panels: The emissivity, $\varepsilon_{22}$, for various Urca pairs normalized to the mass fraction of the parent nucleus.  Right  panels: The 
    luminosity,$L_{34}$, per angular bin for a bin size of $\pi/200$ rad for each Urca pair.  
    The top, middle, and bottom rows correspond to a field B of 10$^{14}$ G, 10$^{15}$ G, and 10$^{16}$ G respectively.}
    \label{rho_em_lum}
\end{figure}

For example, the crusts and oceans of neutron stars have been modeled for 
various dipole fields defined by the field at the NS polar region.  The density
at which various Urca pairs are viable for various polar fields
is shown in Figure \ref{rho_em_lum} as a function of polar angle on the NS 
surface.  For an assumed dipole field, the field is dependent on the NS polar angle
at the surface.  A 12 km diameter star with M=1.4 M$_\odot$ was considered.  A 
constant temperature and $Y_e$ of $T_9=0.51$ and $Y_e$=0.41 are assumed in each
case.  For these fixed parameters and a fixed dipole field, the density
at which $\mu_e\sim Q$ is found.  This is plotted in the left column of 
Figure \ref{rho_em_lum} for fields of $B(\theta=0)$ of 10$^{14}$ G, 10$^{15}$ G,
and 10$^{16}$ G.  Because the field is not constant across the NS surface, the
electron chemical potential is not constant for constant density.

Using the densities computed in Figure \ref{rho_em_lum}  (left panels), the neutrino
emissivity as a function of polar angle in the NS ocean is plotted in the 
same figure (center).  Here, the emissivity is normalized to the
mass fraction of the relevant nuclei in these panels, $\varepsilon/X$.  The 
emissivity is computed using the approximation of \cite{deibel16}.  Because
we are interested in bulk behaviors of the neutrino emissivity
and luminosity, we adopt the approximate weak rates of \cite{arcones10} using the phase
space extracted from Equation \ref{derv}~\citep{famiano20}.  However,
given the extraction of the $ft$ values from the rates presumed in \cite{deibel16}, 
the rate ratios given by the phase space differences with and without the magnetic
fields, $\lambda(B\ne 0)/\lambda(B=0)$ are expected to be independent of any nuclear structure effects in this evaluation.

The luminosity, $L_{32}$
is shown on the right side of Figure \ref{rho_em_lum} assuming an axisymmetric
field for polar angle increments of $\Delta\theta=\pi/200$ radians.  The radial
thickness of each zone is calculated using the formulation of
\cite{schatz14}.  For this calculation, a temperature of T$_9$=0.51, an electron
fraction of $Y_e=0.41$, a crust radius of 12 km, and a local gravity of 
$g=1.85\times 10^{14}$ cm s$^{-2}$ are assumed.  The smaller overall solid angle near
the poles results in an overall reduction of the total luminosity, though
the emissivity may be larger in this region.  The value of the neutrino emissivity, $\varepsilon$, is 
proportional to the weak interaction rates and is computed following the prescription of 
\citep{deibel16}.

There are multiple effects contributing to the behavior of the neutrino luminosity as
a function of field.  At the densities necessary for an Urca pair to exist, only the 
high(low)-energy tail of the electron energy spectrum is relevant for $\beta^-$(EC) decay. 
Also, a change in the local magnetic field will shift the overall electron chemical
potential as shown in Figure \ref{mu_theta_fig}.  In this case, shifting the electron chemical potential in one direction or the other will result in a shift in the electron
phase space, resulting in $\beta$ decays or electron captures being impossible.  As the magnetic field increases, the overlap between the Landau level distribution, the Fermi
distribution, and the electron momentum space can interfere constructively or destructively,
resulting in dramatic shifts in the optimum density at which an Urca pair can exist and 
changing the overall emissivities.  This will be discussed below.

In this figure  one can see that, for very high fields ($\sqrt{eB}\gtrsim$ Q), the entire electron energy 
spectrum
is confined to a single Landau level.  As the field increases near the poles,  the electron
chemical potential decreases, resulting in a lower density at which an Urca pair can 
exist.  This can result in a reduced emissivity.  In addition, as the
Landau level spacing increases, and the tail of the lowest Landau level  ($E=m_e $)
is more prominent in the electron energy spectrum.  As the field decreases, and the Landau
levels move closer together (but with a level spacing still greater than the Q value), the
tail of the lowest Landau level becomes less prominent in the electron spectrum,
resulting in a decrease in the overall rates.  

For a very low field, $\sqrt{eB}\ll Q$,  the effects of the Landau level spacing on the 
overall electron spectrum are minimal.  Here, any change in field along the surface of the NS has little or no effect on the electron chemical potential, and the optimal density for
an Urca pair does not change significantly.  

The ``intermediate field'' regime ($\sqrt{eB}\sim Q$) is particularly interesting as the 
interference between the Landau level distribution and the Fermi distribution becomes 
prominent.  In this regime, a small shift in the magnetic field can result in a 
shift in the optimum density of an Urca pair through a shift in the electron chemical 
potential.  This contributes to a shift in the emissivity.  Because the location of Landau 
levels strongly  affects  the availability of electrons to decay or capture,
the effect can be magnified.
\begin{figure}
    \gridline{
        \fig{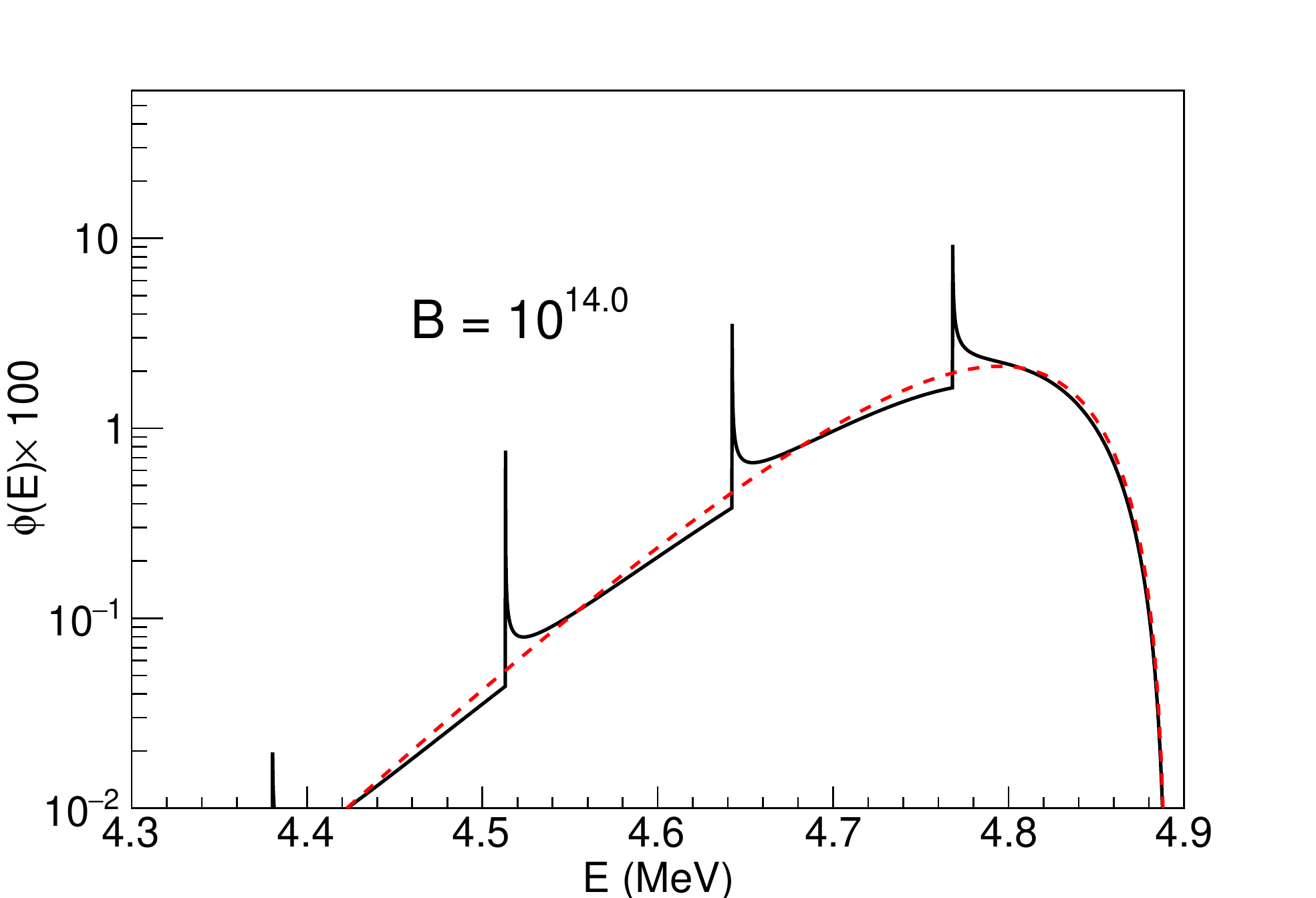}{0.25\textwidth}{}
        \fig{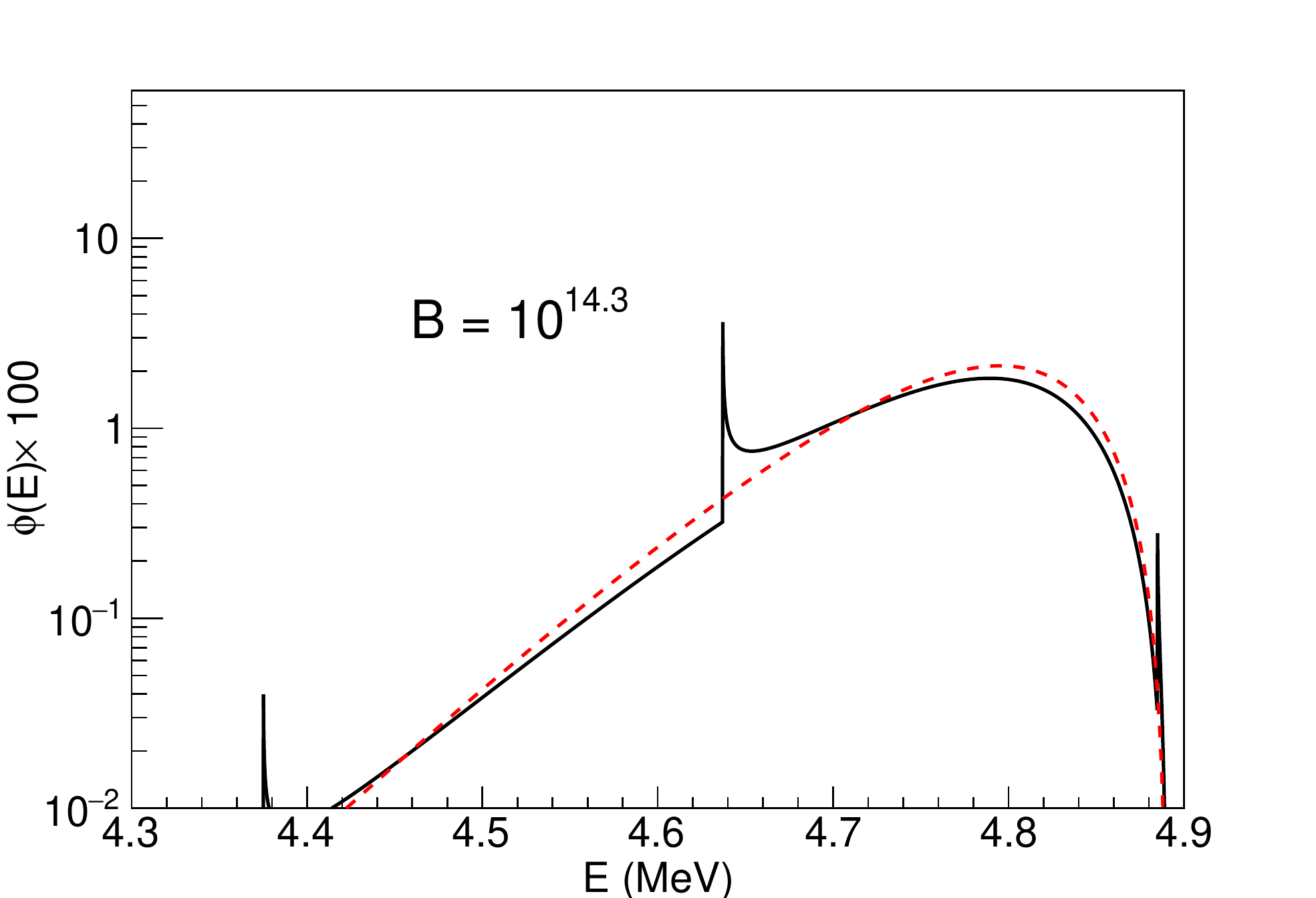}{0.25\textwidth}{}
        \fig{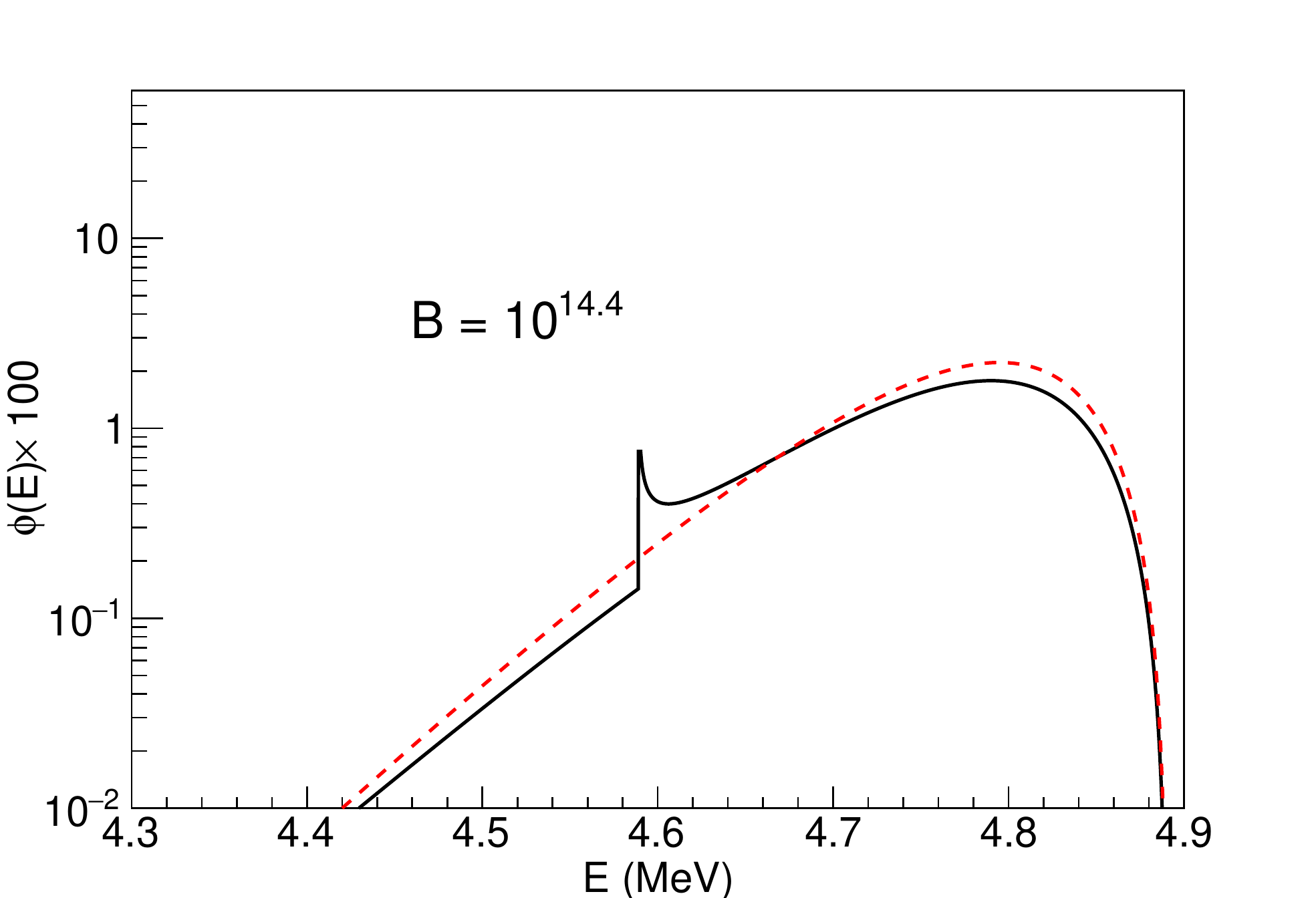}{0.25\textwidth}{}
        \fig{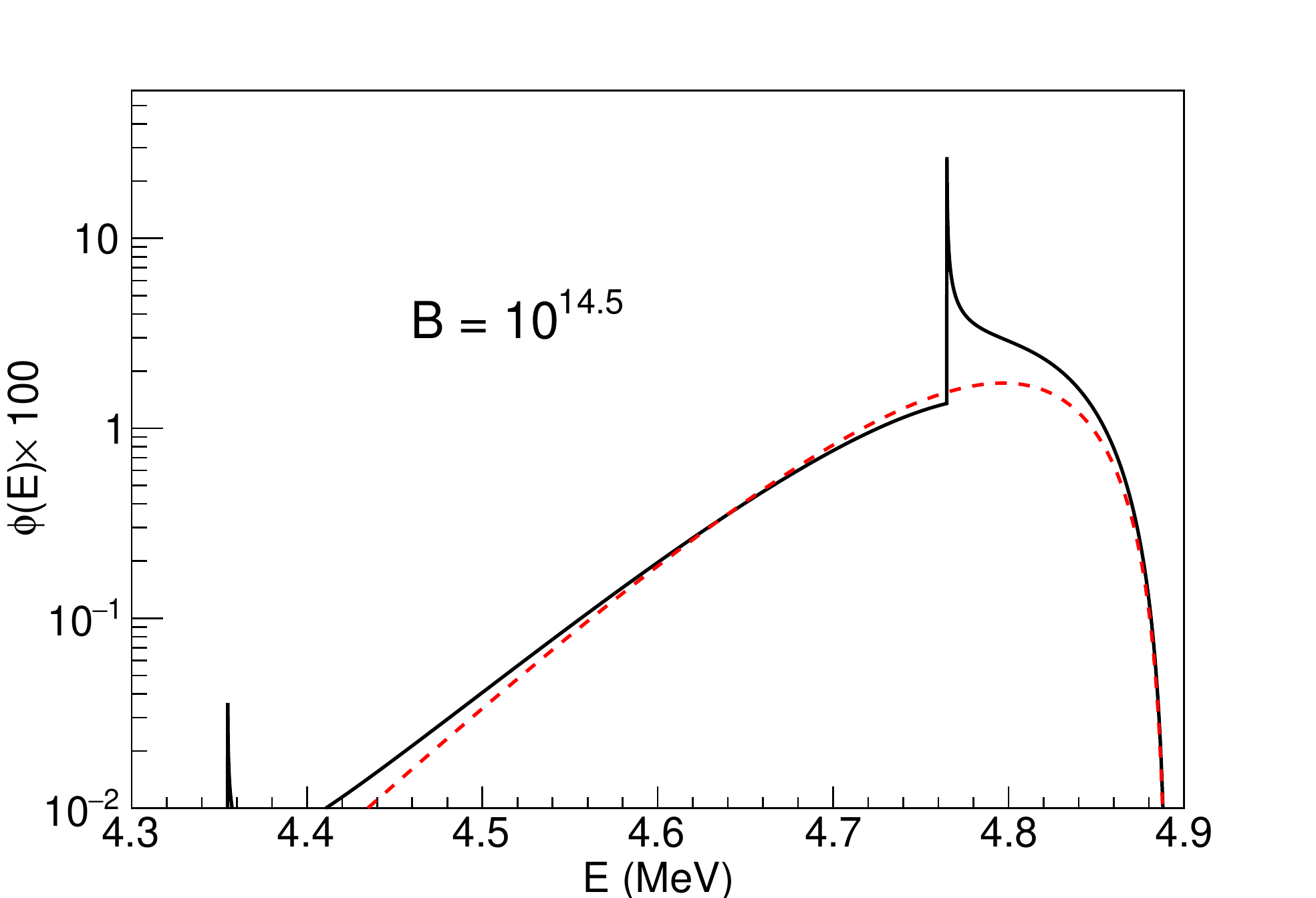}{0.25\textwidth}{}
    }
    \gridline{
        \fig{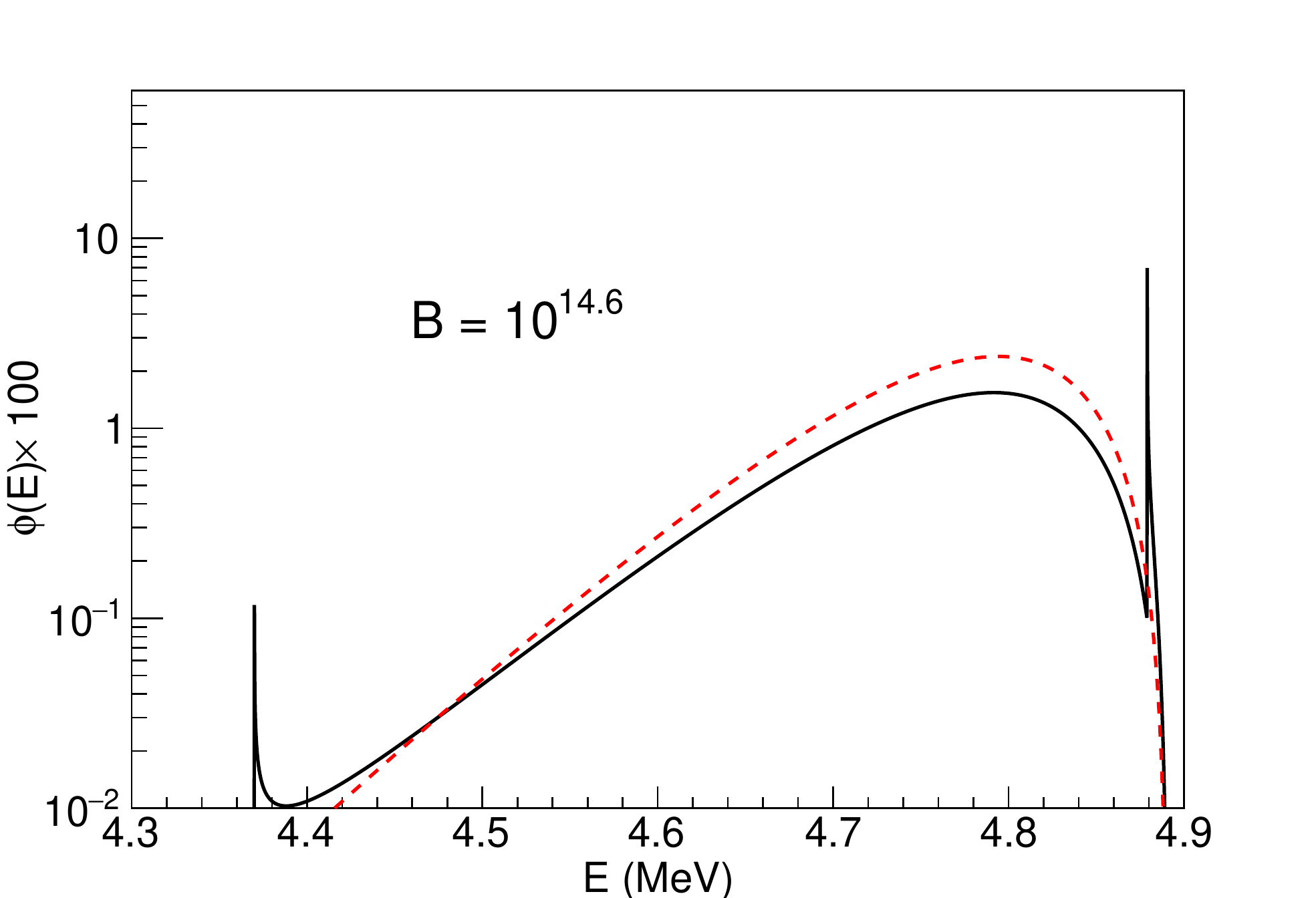}{0.25\textwidth}{}
        \fig{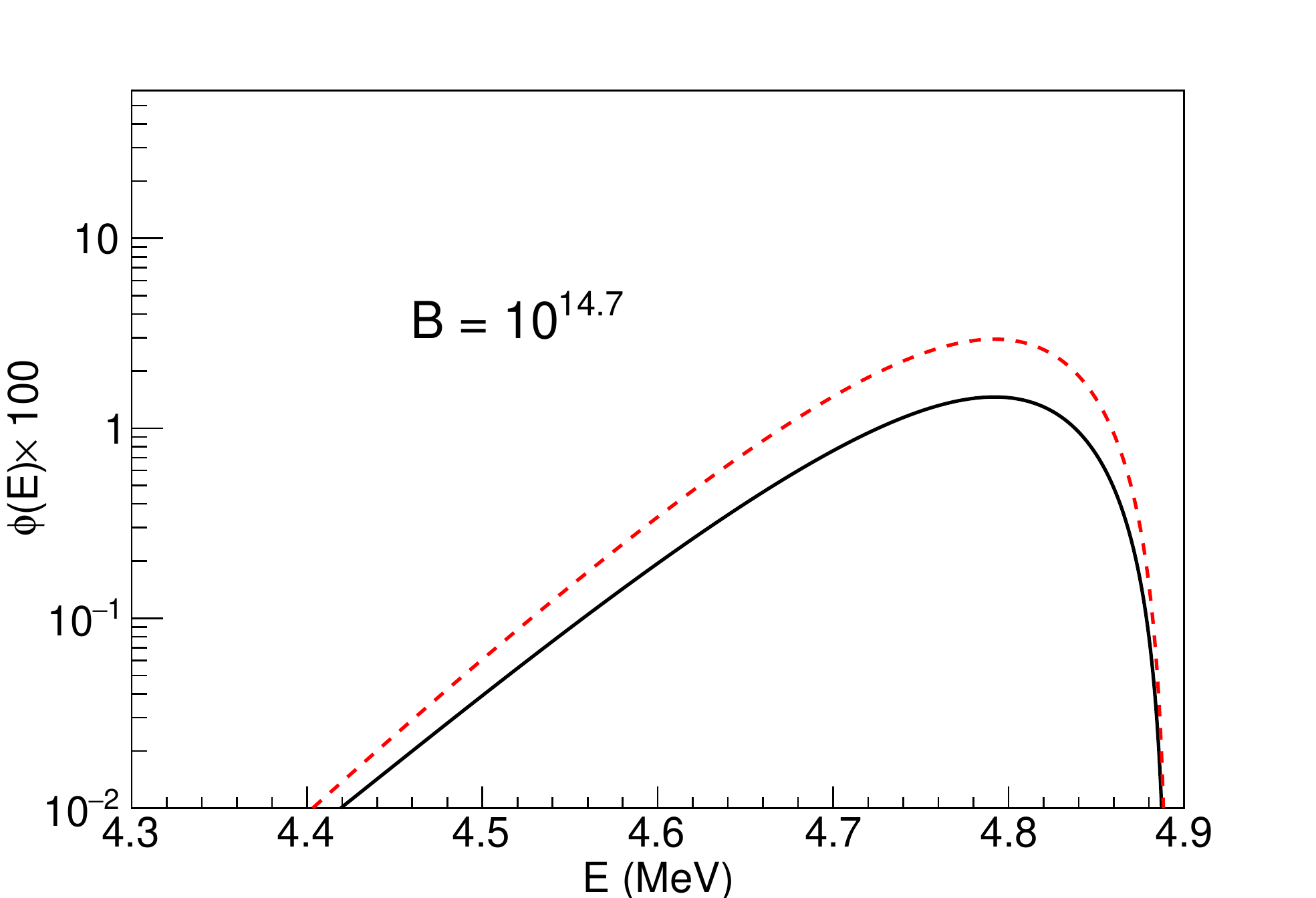}{0.25\textwidth}{}
        \fig{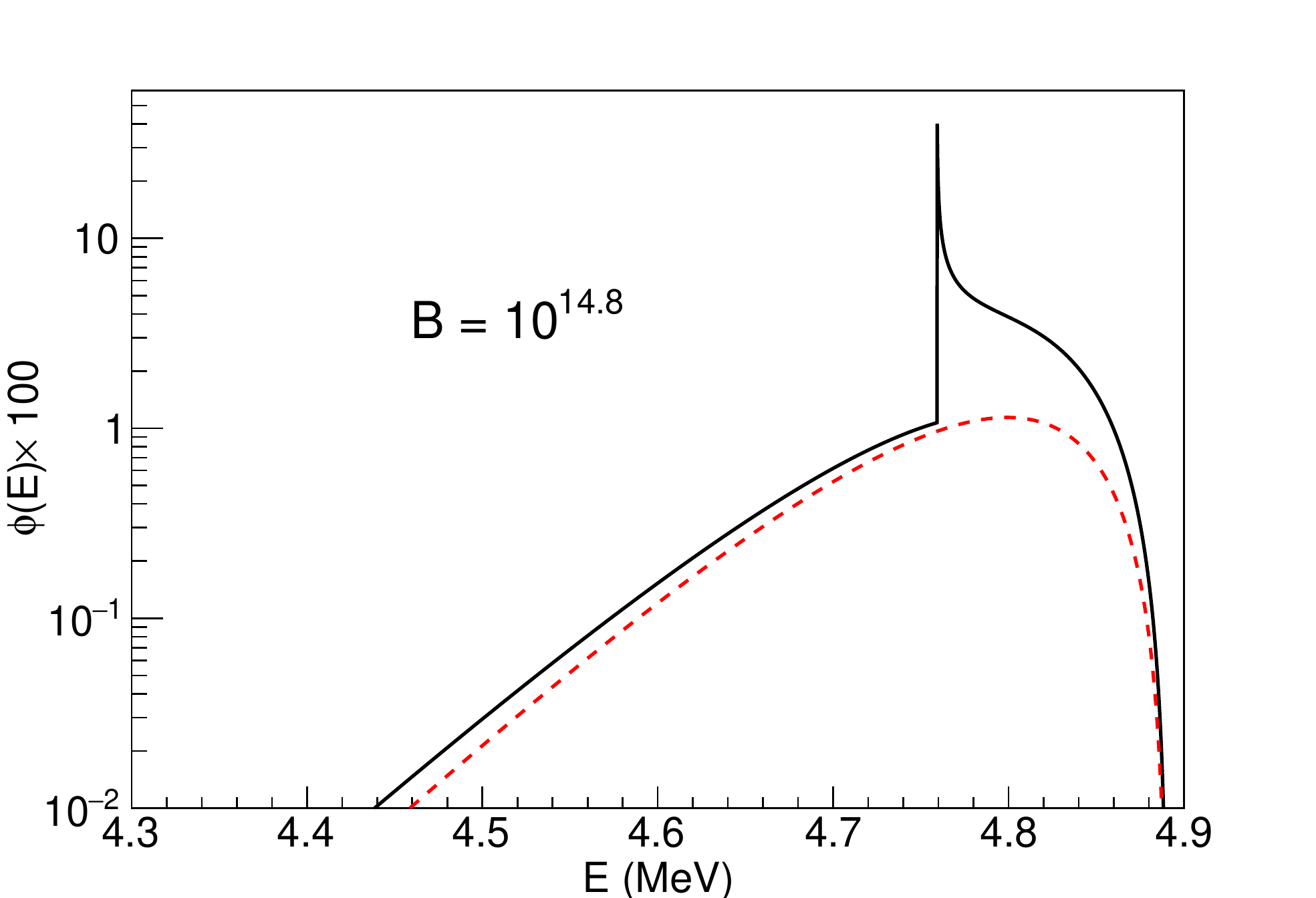}{0.25\textwidth}{}
        \fig{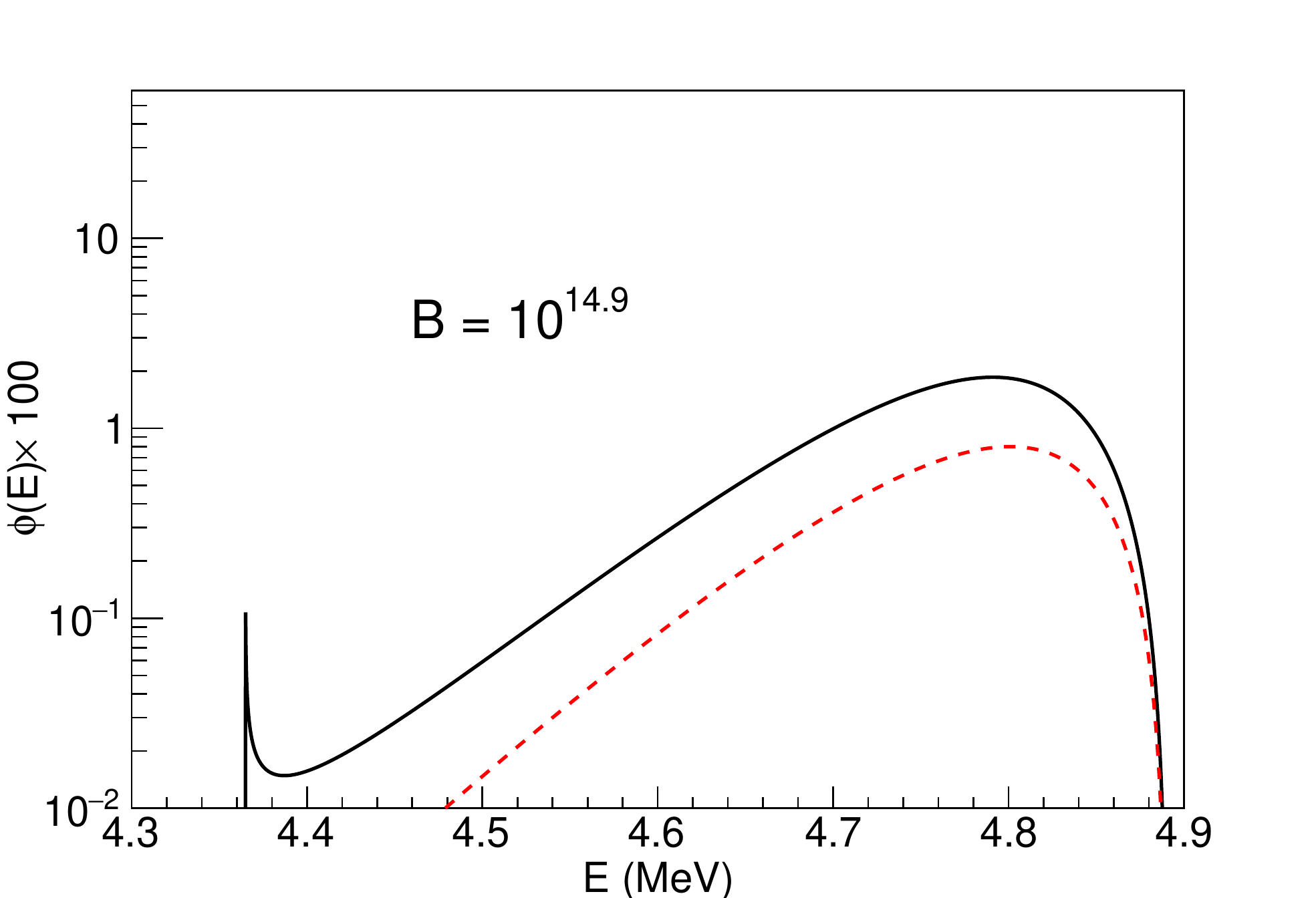}{0.25\textwidth}{}
    }
    \caption{$\beta^{-}$ Decay energy spectrum for various magnetic fields (G) for the $^{23}$Ne decay.  Red dashed lines correspond to the spectrum for B=0, and the solid black lines correspond to the spectrum for the fields indicated in each panel. Each plot in this figure is calculated at T$_9$=0.51, $Y_e$=0.41, and a density at which $\mu_e=Q_\beta$.}
\label{integrand_B}
\end{figure}

This can be seen in Figure \ref{integrand_B}, which shows the electron phase space distribution
for $^{23}$Ne decay in the $^{23}$Na$\leftrightarrow^{23}$Ne Urca pair.  In each
panel in this figure, the optimum density for this Urca pair to exist is computed (i.e., $\mu_e=Q_\beta$), and
the resulting electron phase space spectrum (the integrand in the classical decay rate
integral) is shown.

For a low field of 10$^{14}$ G, the Landau level spacing is slightly less than the
energy range of available electrons.  As the field changes, the Landau level spacing
changes, as do the positions of individual Landau levels.  One sees that as the
field increases  above  10$^{14}$ G, Landau levels can change location with respect to the
edge of the Fermi distribution.  At 10$^{14.5}$ G, a Landau level exists at the right
edge of the Fermi distribution with a less-prominent contribution from a level near 4.4 MeV.  Most of the energy spectrum is dominated by 
the tail of the lower-energy Landau level.  At a field of 10$^{14.7}$ G, the Landau level spacing is
larger than the available phase  space created by the Fermi distribution and the electron momentum 
distribution. For this reason, much of the available electron energy spectrum is dominated by the tail
of a low-energy Landau level.
As the field increases towards 10$^{15}$ G, the degeneracy of available Landau levels increases as electrons
are placed in only the lowest few Landau levels, and the occupancy of the tails of the distributions becomes
more important in the electron energy spectrum.   
When the field is high enough such that only the lowest Landau level is occupied, the tail of the distribution becomes quite prominent, and the electron energy spectrum is dominated by the single Landau
level.  

\section{Conclusions}
We explored the dependence and evolution of Urca pairs in crusts and oceans of magnetized
neutron stars.  Because the electron
chemical potential depends on the environmental magnetic field, the presence or absence of
Urca pairs must also  depend on the external field.  Urca pairs which may exist at a certain
temperature and density, $\rho Y_e$, at zero magnetic field may not exist for a non-zero field.  
Conversely, Urca pairs which do not exist at specific temperatures and  densities may appear in
the presence of an external field.  The presence of a magnetic field may also make excited states
available in EC-$\beta^-$ pairs.  In this case, two nuclei which may not transition to low-lying
states, may undergo transitions to excited states in the presence of an external field.

We also explored the evolution of Urca pairs along the surface of a magnetized neutron star.
While we have assumed a dipole field for simplicity, this is sufficient to convey the concept that 
the variations in the field on the surface of a NS may change Urca pair locations within the
crust/ocean of the star.  Here, we have calculated the density at which an Urca pair may exist
in a NS crust/ocean for a constant crust temperature of $T_9 = 0.51$.  Of course, we note that the
temperature of the crust  varies with depth, so we acknowledge the simplification 
in this evaluation.  This variation in density will result in a change in the neutrino
emissivity and subsequent luminosity as a function of polar angle.  We find that the emissivity 
(normalized to the mass fraction of the Urca pair studied) subsequently depends on the NS latitude,
resulting in  changes in the luminosity as a function of latitude.  

However, because the density at which an Urca pair may exist  in a NS  is not constant, this 
ultimately affects the actual presence of an Urca pair at a location  in a NS.
This is because the density and the mass fraction changes with depth in the NS crust. At one
location on the NS surface, an Urca pair may exist at one density while it exists at a different
density at another location  in the NS.  However, the mass fractions of the Urca pair
also vary at different densities/radii.  For this reason, the emissivity is presented as
normalized to  the mass fraction, and a true computation of the emissivity must be multiplied by the 
mass fraction for the radius at which a specific density exists.  Figure \ref{rho_em_lum} can
be used to scale Urca pair emissivity as a function of latitude for a known mass fraction, X.
Thus, the emissivity can be expressed as a functional, 
$\epsilon = \epsilon(T,\rho Y_e, B, X) = \epsilon(T, \rho(r,\theta) Y_e(r), B(\theta))=
\epsilon(T,r,\theta, B_0)$, where $B_0$ is the magnetic field at the NS pole.  

It may very well be possible that the mass fraction in the region at which an Urca pair may be
viable is zero, while it is non-zero in another viable region.  Thus,  an Urca pair may
exist in one part of the NS, but not in another.   We have shown  that the addition of a
magnetic field to NS luminosity calculations adds another layer of complexity to the total
emissivity calculation, and ultimately may result in uneven NS cooling along the surface of
the star.

There may be several ramifications of this result which will be explored in subsequent papers. 
First, we find that the bulk luminosity of
a NS is dependent on its magnetic field.  This can ultimately change the cooling curve of the star.  
Perhaps variations in NS luminosities are a result of the surface field.  Likewise, limits can
be placed on NS surface fields from observations of their cooling.

Second, uneven neutrino emissivity on the NS surface can result in uneven cooling and heating in the 
crust.  This can have affect the thermal equation of state on the surface and may be a way of explaining
NS crust quakes or possibly even NS kicks.

Certainly,  an exhaustive treatment of the overall complexity of this problem is  beyond the scope of the present work.   More
precise work is needed.  In particular, a thorough evaluation of the presence or absence of Urca pairs as a
function of density, temperature, and magnetic field is needed.  From this, luminosity maps can be
produced for NSs with various field configurations.  In order to produce these, realistic models of 
ocean/crust abundances can be developed.   While these will be explored in  future work,  our current results introduce the possibility that magnetized neutron
stars may have uneven cooling as a result of variations in the interior magnetic field, and the overall bulk luminosity
changes with the field configuration and magnitude.
\bibliography{sample631}{}
\bibliographystyle{aasjournal}


\begin{acknowledgements}
T.K. is supported in part by Grants-in-Aid for Scientific Research of JSPS (17K05459, 20K03958).  A.B.B. is supported in part by the U.S. National Science Foundation Grants No. PHY-2020275 and PHY-2108339.  M.A.F. is supported by National Science Foundation Grant No. PHY-1712832 and by NASA Grant No. 80NSSC20K0498. K.M. is supported by Research Institute of Stellar Explosive Phenomena at Fukuoka University and JSPS KAKENHI Grant Number JP21K20369.  M.A.F., G.J.M.  and A.B.B. acknowledge support from the NAOJ Visiting Professor program.  Work at the University of Notre Dame (G.J.M.) supported by DOE nuclear theory grant DE-FG02-95-ER40934.
\end{acknowledgements}
\end{document}